\documentclass[epj
]{svjour}

\usepackage{graphics}
\usepackage{epsfig}
\usepackage{subfigure}
\usepackage{mathptmx}
\usepackage{amsfonts}
\usepackage{amsmath}
\usepackage{amssymb,bm}
\usepackage{bbm}
\usepackage{bigstrut}
\usepackage{fix-cm}
\usepackage{array}
\usepackage{widetext}
\usepackage{enumerate}
\usepackage[usenames,dvipsnames]{color}
\usepackage[hidelinks]{hyperref}
\usepackage[T1]{fontenc}
\usepackage{url}

\newcommand{\cc}[1]{\text{\textbf{\textit{#1}}}}

\newcommand{\ket}[1]{\vert #1 \rangle}

\renewcommand{\jmath}{j}

\begin{document}
\title{From the liquid drop model to lattice QCD}
\subtitle{A brief history of nuclear interactions}
\author{Vittorio Som\`a
}               % Do not remove
\institute{
IRFU, CEA, Universit{\'e} Paris-Saclay, 91191 Gif-sur-Yvette, France
}

\date{\today}

\abstract{The present article aims to give a concise account of the main developments in nuclear structure theory, from its origin in the 1930s to date, taking the modelling of inter-nucleon interactions as guideline.
}

\maketitle

%%%%%%%%%%%%%%%%%%%%%
\section{Introduction}
\label{sec_intro}
%%%%%%%%%%%%%%%%%%%%%

Mendeleev published his periodic table of elements in 1869.
The first atomic nucleus, $^{1}$H, was discovered more than a hundred years ago. 
Nuclear physics, as we conceive it today, was born in the 1930s with the discovery of the neutron and the theory of weak interactions.
Hence, it is one of the oldest subfields of contemporary physics.
Nevertheless, our understanding of nuclear systems is far from being complete, and several crucial questions concerning the origin, the limits of existence and the properties of nuclei are still unanswered.

A key ingredient in our theoretical description of nuclei is constituted by the basic interactions between their constituents, protons and neutrons.
In spite of eighty years of developments, this remains the most uncertain piece of the nuclear puzzle. 
This article aims to give a concise account of such developments as well as the challenges that can be envisaged for the future.
In parallel, the main many-body models and theories developed in the context of low-energy nuclear physics will be briefly discussed.

%%%%%%%%%%%%%%%%%%%%%
\section{Basic facts and questions about nuclei}
\label{sec_basic}
%%%%%%%%%%%%%%%%%%%%%

Atomic nuclei constitute the (positively charged) central cores of atoms, are made of protons and neutrons and contain nearly all the atomic mass.
Each chemical element is specified by the number of protons in the nucleus of its atoms, $Z$, also known as the atomic number. 
For each $Z$, nuclei with different numbers of neutrons $N$ can exist, and are called isotopes\footnote{Analogously, nuclei with the same neutron number $N$ but different proton number $Z$ are referred to as isotones.}. 
Here are some of the basic questions about these systems.

\begin{itemize}
\item[$\bullet$] \textit{How many nuclei exist?}\\
Actually, what do we mean here by \textit{exist}? 
A given combination of protons and neutrons ($Z$, $N$) exists as a nucleus if they can form at least one state bound with respect to the strong force. 
Quantum states of bound nuclei can be unstable or stable, depending on whether, respectively, they do or do not decay autonomously into other nuclear systems due to the effect of other forces, e.g. electroweak interactions.

As of today we know of 253 
stable and about 3100 unstable isotopes~\cite{Huang17}, the majority of which do not occur in nature but have been synthesised in laboratories on Earth.
All together, they can be displayed as a function of their proton and neutron numbers in the so-called Segr\`e chart of nuclides, see Fig.~\ref{fig_segre}.
In spite of remarkable experimental and theoretical progress of the last decades, not all of the possible bound combinations of $Z$ and $N$ have been identified so far. Depending on the theoretical prediction, from $\sim6000$ to $\sim9000$ isotopes are believed to exist.\\
\item[$\bullet$] \textit{Where are proton and neutron drip lines?}\\ 
\textit{Drip lines} mark the limits of existence of nuclei in the Segr\`e chart. 
Let us start from a given bound and stable nucleus with ($Z$, $N$). 
Let us now add (or remove) neutrons, i.e. move horizontally in the Segr\`e chart.
As we move away from the region of stable isotopes, nuclei eventually become less and less bound and at one point become unbound, i.e. they do not exist anymore.
Such limits (of existence) are respectively the neutron and proton drip lines. 
While the latter has been determined experimentally for more than half of the known elements, the former is established only for $Z \leq 8$.
\\
\item[$\bullet$] \textit{Which is the heaviest possible element?}\\ 
What happens if we keep creating heavier and heavier nuclei? 
Is there a limit, or can we proceed indefinitely? 
At present, the heaviest synthesised element has 118 protons. 
On the other hand, the heaviest stable nucleus we know is $^{208}$Pb with 82 protons.
As we produce heavier and heavier nuclei, will we encounter a new region of super-heavy stable isotopes?
\\
\begin{widetext}
\begin{figure}[b]
\begin{center}
\includegraphics[width=18cm]{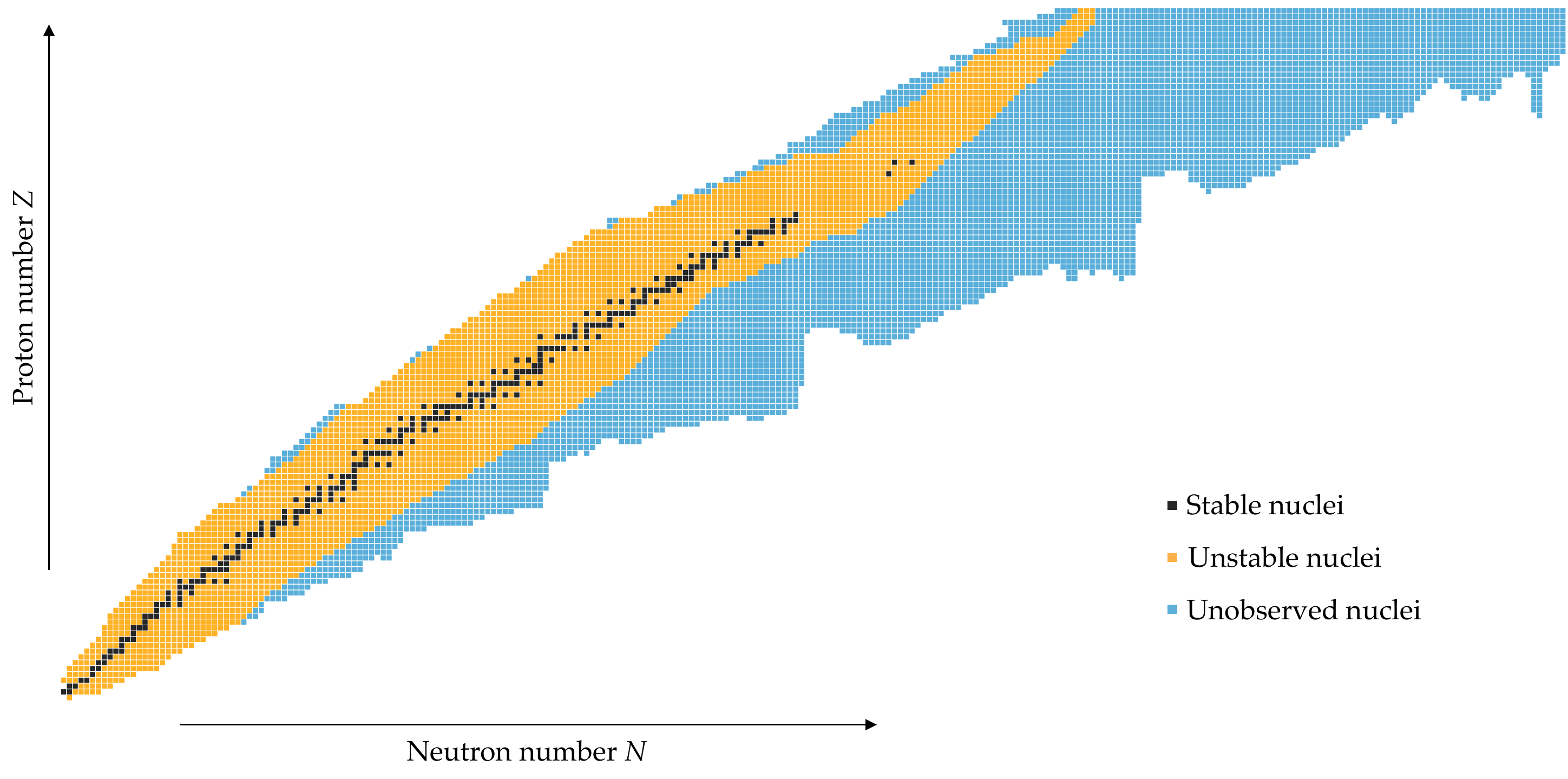}
\end{center}
\caption{Segr\`e chart of nuclides, in which existing isotopes are displayed as a function of their number of neutrons, $N$, and protons, $Z$. Different colours are used to distinguish stable, unstable and unobserved nuclei. The latter are based on the prediction of a given theoretical model.
Adapted from~\cite{BallyChart}.
% + add reference
}
\label{fig_segre}
\end{figure}
\end{widetext}
\item[$\bullet$] \textit{Are magic numbers the same for unstable nuclei?}\\ 
Within a given mass region, we find that some nuclei are significantly more stable than their immediate neighbours\footnote{It means that their total binding energy represents a local maximum in that mass region.}. 
Interestingly, such nuclei are characterised by certain recurring ``magic" numbers of protons and/or neutrons. 
However, as more and more unstable, neutron-rich isotopes were produced and studied, it was noticed that away from the region of stable nuclei (the so-called \textit{valley of stability}), some of these numbers do not correspond anymore to the most stable configurations.
In other words, magic numbers evolve across the Segr\`e chart. How, exactly?
\\
\item[$\bullet$] \textit{How (and where) have known elements been produced?}\\ 
We know that light and medium-mass elements have originated from either Big Bang or stellar nucleosynthesis. However, it not yet clear how and where (about half of the) elements heavier than iron have been produced in the universe\footnote{In 2001, this was included among the \textit{11 greatest unanswered questions of physics} by US National Research Council~\cite{11questions}.}.
As of today, neutron-star mergers and type-II supernova events are the most plausible candidates for inducing the synthesis of these nuclei.
\end{itemize}
These represent some of the most basic open questions regarding nuclear systems.
They mainly concern properties of nuclei in their ground state, in particular the total binding energy associated to a given combination of protons and neutrons.
Nonetheless, atomic nuclei are characterised by several other types of observables and processes, e.g.
\begin{itemize}
\item[$\circ$] Among ground-state properties: in addition to total binding energies, radii (i.e. sizes), shapes, superfluidity, ...\\
\item[$\circ$] Various types of radioactive decays: $\beta$, double-$\beta$, $\alpha$, proton, two-proton, ...\\
\item[$\circ$] Spectroscopy, i.e. the study of excited modes\\
\item[$\circ$] Exotic structures such as clusters and halos\\
\item[$\circ$] Reaction processes: fusion, transfer, knock-out, ...\\
\item[$\circ$] Electro-weak processes\\
\end{itemize}
\vspace{-0.3cm}
The goal of nuclear physicists is to provide a global account and understanding of this rich diversity of phenomena. 
Nuclear experiments aim at either more precise and systematic measurements of these properties, or at extending our knowledge further away from the valley of stability or to novel types of observables.
Nuclear theory aims at developing global and predictive approaches, possibly related to the underlying theory of quantum chromodynamics (QCD).
Both tasks are daunting. 
On the one hand, experiments have to deal with physical systems that are unstable and, in many cases, short-lived. 
On the other hand, nuclei are mesoscopic, complex systems that require an articulated theoretical description.
First, they are quantum mechanical objects.
Second, several scales are at play, from proton and neutron momenta (of the order of $10^8$ eV) going to nuclear separation energies (around $10^7$ eV) and vibrational excitations (around $10^6$ eV) down to rotational excitations (as low as $10^4$ eV).
Third, if neutrons and protons (referred to collectively as nucleons) are chosen as degrees of freedom, most nuclei have too many constituents to be treated exactly but not enough to be treated within statistical mechanics. 
Among others, these represent some of the difficulties that we encounter in nuclear physics and that keep the basic questions discussed above open.

%%%%%%%%%%%%%%%%%%%%%
\section{The nuclear many-body problem}
\label{sec_manybody}
%%%%%%%%%%%%%%%%%%%%%

\subsection{Liquid drop model}
\label{sec_liquid_drop}
\begin{figure}[t]
\begin{center}
\includegraphics[width=8.5cm]{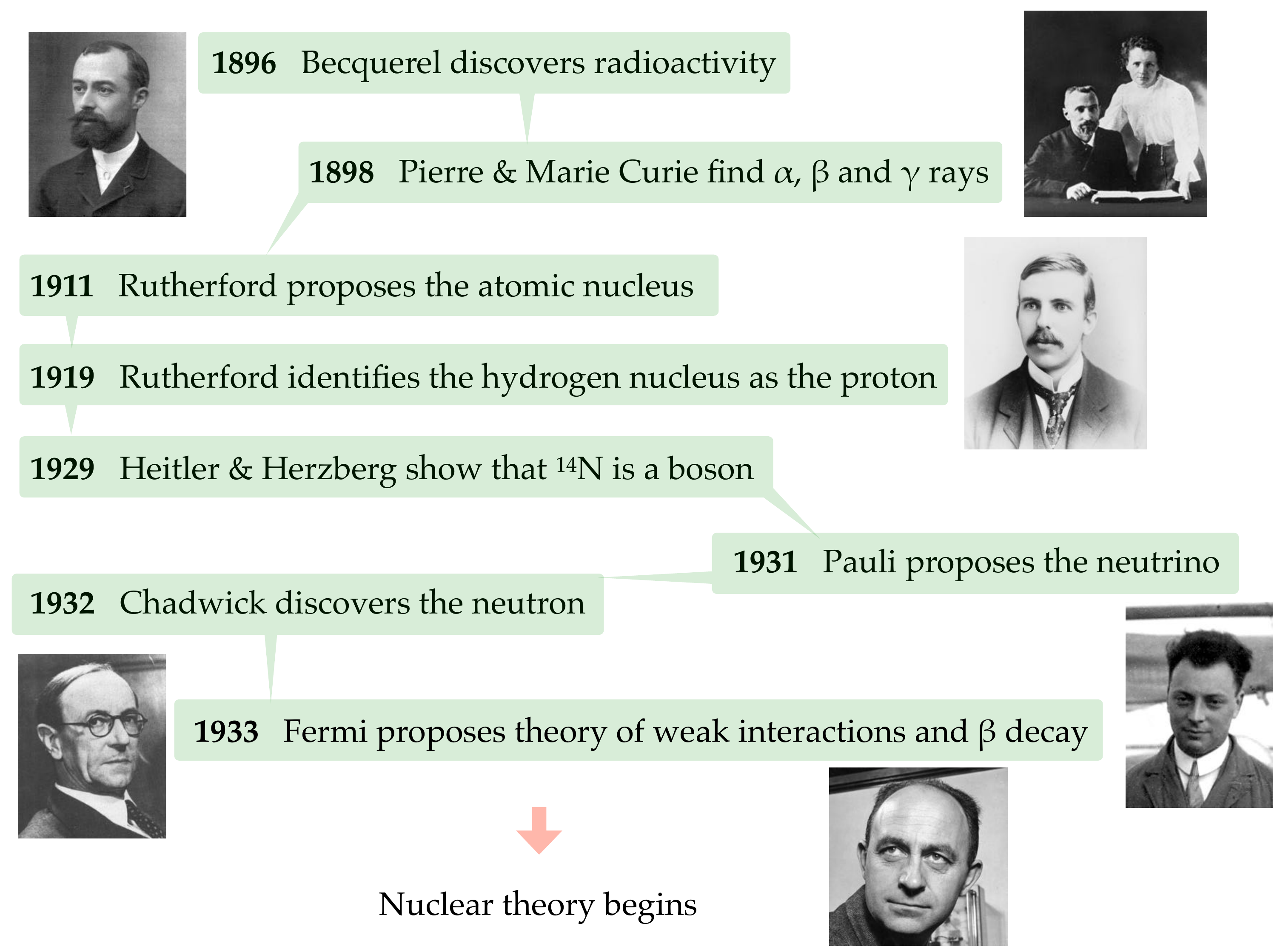}
\end{center}
\caption{Main discoveries at the basis of nuclear physics.}
\label{fig_historical}
\end{figure}
Historically, nuclear theory can be set to begin in the 1930s following a series of milestones that starts with the discovery of radioactivity by Becquerel in 1896 and ends with the theory of weak interactions by Fermi in 1933, see Fig.~\ref{fig_historical}.
The first models, developed among others by Gamow, Bohr and Wheeler, pictured the atomic nucleus as a (suspended) drop of incompressible liquid with a surface tension. 
This idea led Weizs{\"a}cker and Bethe, in 1935, to conceive a semi-empirical formula for the total energy $E$ of a nucleus where nuclear binding emerges from different competing processes~\cite{Weizsacker35},
\begin{equation}
E(Z,N) = a_v~A - a_s~A^{2/3} -a_c \frac{Z^2}{A^{1/3}} - a_a \frac{(N-Z)^2}{4 A} -\frac{\delta}{A^{1/2}} \; ,
\label{eq_semi}
\end{equation}
$A=Z+N$ being the total number of nucleons, i.e. the mass number.
The various terms are associated to different parameters accounting for effects of volume ($a_v$), surface ($a_s$), Coulomb force between protons ($a_c$), neutron-proton asymmetry ($a_a$) and pairing ($\delta$).
Refined over the years, the Weizs{\"a}cker-Bethe semi-empirical formula results successful in explaining the global trend in binding energies of stable nuclei.
Nevertheless, such a simplistic semi-classical model can not correctly describe fine features nor other nuclear properties starting with the excitation modes. For that, and more, a full quantum mechanical treatment is needed.

\subsection{Which degrees of freedom?}

In general, the choice of the basic constituents of a model is crucial for its successful description of physical phenomena within a given energy domain.
In the case of atomic nuclei, the most obvious degrees of freedom are protons and neutrons. 
As a result, one has to solve a many-body quantum-mechanical problem, typically in the form of a Schr{\"o}dinger equation. 

Even if protons and neutrons appear to be the most natural degrees of freedom, the wide range of energy scales associated to nuclear observables already suggests that they might not always be the most suitable ones. 
For instance, it will be extremely complicated to describe low-energy collective excitations in heavy nuclei in terms of individual nucleons, whereas using quantised collective modes might be more efficient.
On the other hand one may object that protons and neutrons are not fundamental building blocks, and one should start instead from QCD degrees of freedom.
Indeed, a full QCD treatment of bound states of baryons is becoming feasible nowadays thanks to large-scale lattice simulations. 
A qualitative description of light nuclei is at reach and quantitative results are expected in the near future. 
In the present article I will only briefly touch upon these alternatives and rather focus on the modelling of nuclei in terms of protons and neutrons.

\subsection{Nuclear structure, matter and reactions}

The large majority of theoretical approaches to nuclear systems, therefore, uses nucleonic degrees of freedom. 
Most of them address the properties of an isolated nucleus, i.e. a system with fixed numbers of protons and neutrons, in its ground state or in some low-lying excited state. The relevant equation to describe such a system is then the time-independent Schr{\"o}dinger equation
\begin{equation}
H \ket{\Psi^A_k} = E^A_k  \ket{\Psi^A_k} \; ,
\label{eq_schroedinger}
\end{equation}
an eigenvalue equation yielding $\ket{\Psi^A_k}$, the many-body wave function corresponding to the excited state $k$, and $E^A_k$, its associated energy.
$H$ is the nuclear Hamiltonian, which encodes interactions between nucleons and is independent of $A$, i.e. the same for all nuclei\footnote{To be precise, an $A$-dependent term is usually added to correct for the contamination induced by center-of-mass (c.o.m.) motion.
Since we are interested in the intrinsic excitations of the system, we typically subtract from the total $H$ the nucleus-dependent c.o.m. kinetic energy $T_{c.o.m.}(A)$ and end up working with $H_{int} (A) = H - T_{c.o.m.}(A)$.
Nevertheless, this is a well-defined correction that does not impact the modelling of the $A$-independent $H$.}.
Nuclear structure theory concerns the study of Eq.~\eqref{eq_schroedinger} with $A$ going from two to as high as possible, up to the limits of existence of super-heavy nuclei.

As will be discussed in the following, the solution of the Schr{\"o}dinger equation becomes more and more cumbersome as we increase $A$.
However, a particular case is constituted by the limit\footnote{Strictly speaking I refer to the thermodynamic limit where both the number of particles and the volume $\mathcal{V}$ go to infinity,  $A~\rightarrow~\infty$ and $\mathcal{V}~\rightarrow~\infty$, while the number density $\rho=A/\mathcal{V}$ remains constant.} $A \rightarrow \infty$, where surface effects disappear and the use of a plane-wave basis greatly simplifies the computational steps. 
Such an idealised system is referred to as \textit{nuclear matter} and has represented a useful testing ground for many-body methods and nuclear interactions over the years. 
After the discovery of neutron stars, nuclear matter has actually become less idealised. 
Its study allows modelling the fermionic matter that is thought to form the core of these compact stars.

In reality, we typically gain information on nuclear systems via reactions. 
From the study of a reaction process one should then (carefully) extract details about the structure of a given nucleus. 
To this extent, a rigorous approach would require the use of the time-dependent Schr{\"o}dinger equation
\begin{equation}
H \ket{\Psi^{A+B \rightarrow C+D} (t)} = i \hbar \frac{\partial}{\partial t} \ket{\Psi^{A+B \rightarrow C+D} (t)} \; ,
\label{eq_tds}
\end{equation}
where the wave function of the total system $A+B \rightarrow C+D$ has to be considered\footnote{This is the simplest possibility where two reactants and two reaction products are involved. However, other types of reactions can be envisaged with e.g. more than two final products, particles other than nucleons or nuclei (photons, neutrinos, ...), etc.}.
Here we discuss only approaches to the structure problem, Eq.~\eqref{eq_schroedinger}. 
Nevertheless, it is important to keep in mind that structure observables are often determined via a reaction cross section that should in principle involve Eq.~\eqref{eq_tds}.

\subsection{Effective vs ab initio approaches}
\label{sec_eff_vs_ab}

Let us inspect more closely the many-body Schr{\"o}dinger equation.
While for certain quantum-mechanical systems the Hamiltonian is well defined (e.g. electrons in an atom or a molecule, thanks to our good knowledge of Coulomb interaction), this is not the case for atomic nuclei. 
The first difficulty in solving Eq.~\eqref{eq_schroedinger} thus relates to determining the nuclear Hamiltonian itself.
In principle, inter-nucleon forces arise in QCD as residual interactions between systems of bound quarks (in analogy to van der Waals forces in molecular physics).
However, QCD at low-energies is non perturbative and calculations of many-baryon systems are extremely challenging (see Section~\ref{sec_challenges}). 
This implies that some modelling, more or less phenomenological, has to be invoked in the determination of $H$.

What is its form and what are its properties? 
Let us recall that we consider here the general approach in which structureless protons and neutrons are the relevant degrees of freedom.
The fact that, in reality, they do have an internal structure implies that, in addition to (effective) nucleon-nucleon (NN) interactions, $H$ unavoidably contains three-body, four-body, ... up to $A$-body terms whenever $A$ nucleons are present.
Are all these terms really necessary or can we get away with just pairwise nucleon interactions? 
We will see in Sec.~\ref{sec_3bf} that at least three-body forces are mandatory for a satisfactory description of nuclei and nuclear matter.

In addition to spatial coordinates and momenta, nucleons are described by spin and isospin quantum numbers. How do different operators act in these spaces?  
In Sec.~\ref{sec_properties} we will discuss their basic properties and will see that, e.g., state-of-the-art NN interactions contain up to 15 or 20 different operators.
In any case, it is evident that the modelling of $H$ is not trivial.
\begin{figure}[t]
\begin{center}
\includegraphics[width=9cm]{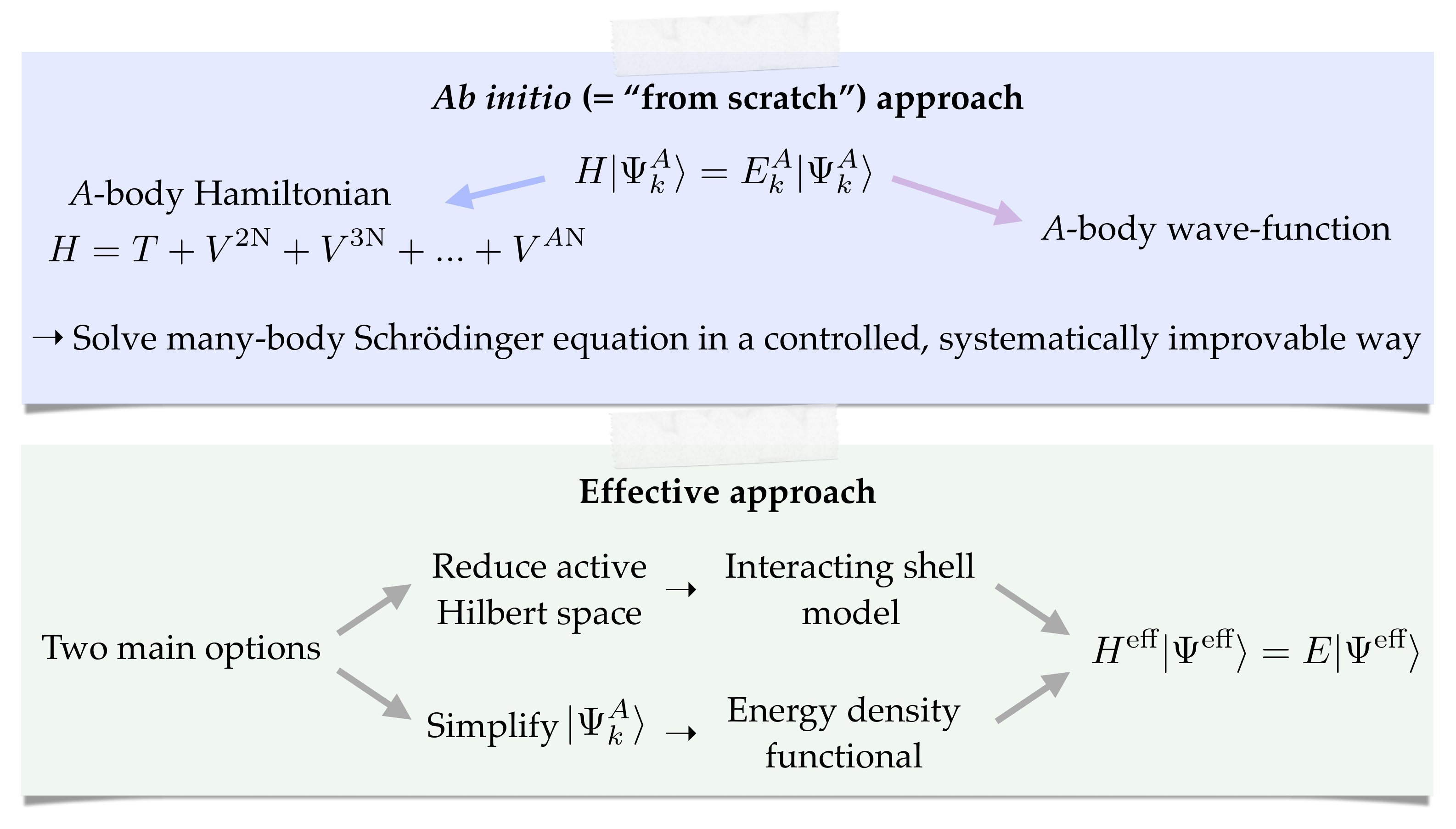}
\end{center}
\caption{Schematic view of ab initio and effective approach to the nuclear many-body problem.}
\label{fig_ab_eff}
\end{figure}
Once the Hamiltonian is at hand, Eq.~\eqref{eq_schroedinger} needs to be solved. 
The $A$-body wave function $\ket{\Psi^A_k}$ is a function of the coordinates, spin and isospin of each nucleon, i.e. of $5 \times A$ variables.
For small values of $A$, exact techniques can be applied. 
As $A$ increases, as will be discussed in the following, the computational cost of these techniques becomes prohibitive. 

The approach described above goes under the name of \textit{ab initio}\footnote{From the latin ``from scratch'' or ``from the beginning'', it indicates approaches that aim to describe a certain physical system from first principles.} and will be discussed, in conjunction with the modelling of the nuclear Hamiltonian itself, in Sections~\ref{sec_old} and \ref{sec_new}.
Its main advantage resides in its predictive character: if one succeeds in building a two-body Hamiltonian that well reproduces the two-body system, a two- plus three-body Hamiltonian that well reproduces the three-body system, and so on, then one hopes to successfully extend it to \textit{any} system.
Even though significant progress has been made in recent years, such calculations remain challenging and the domain of application of ab initio techniques is, at present, limited to $A \sim 100$.

There exist two main alternatives to the ab initio approach, see Fig.~\ref{fig_ab_eff}.
The first option is to reduce the number of active nucleons, thereby reducing the dimensionality of $\ket{\Psi^A_k}$. 
This allows exact diagonalisations of Eq.~\eqref{eq_schroedinger} in reduced one-body Hilbert spaces and, as a result, a wider reach across the Segr\`e chart.
This route is the one followed by the \textit{interacting shell model}.
The second option consists in keeping all $A$ nucleons active but simplifying the $A$-body wave function $\ket{\Psi^A_k}$. 
As a result, the solution of Eq.~\eqref{eq_schroedinger}, at least in a first step, does not require a full and costly diagonalisation but can be obtained via relatively inexpensive many-body techniques.
This is the route followed by \textit{energy density functionals}.
In both cases, the Hamiltonian has to be suitably modified to either include correlations outside the active space (in the case of the interaction shell model) or correlations that would normally be encoded in the full many-body wave function (in the case of energy density functionals).
As a result, one solves
\begin{equation}
H^{\text{eff}} \ket{\Psi_k^{\text{eff}}} = E_k  \ket{\Psi_k^{\text{eff}}} \; 
\label{eq_schroedinger_eff}
\end{equation}
instead of Eq.~\eqref{eq_schroedinger}.
The advantage is a reduced computational cost.
The price to pay is a disconnection from the ab initio Hamiltonian\footnote{In Section~\ref{sec_abinitio} we will see that an ab initio implementation of the interacting shell model has been developed in recent years.}, with subsequent loss of predictive power.
These effective approaches will be discussed in the following Section~\ref{sec_effective}.

%%%%%%%%%%%%%%%%%%%%%
\section{Effective approaches}
\label{sec_effective}
%%%%%%%%%%%%%%%%%%%%%

\subsection{Independent-particle model and mean field}

\begin{figure}[t]
\begin{center}
\includegraphics[width=8cm]{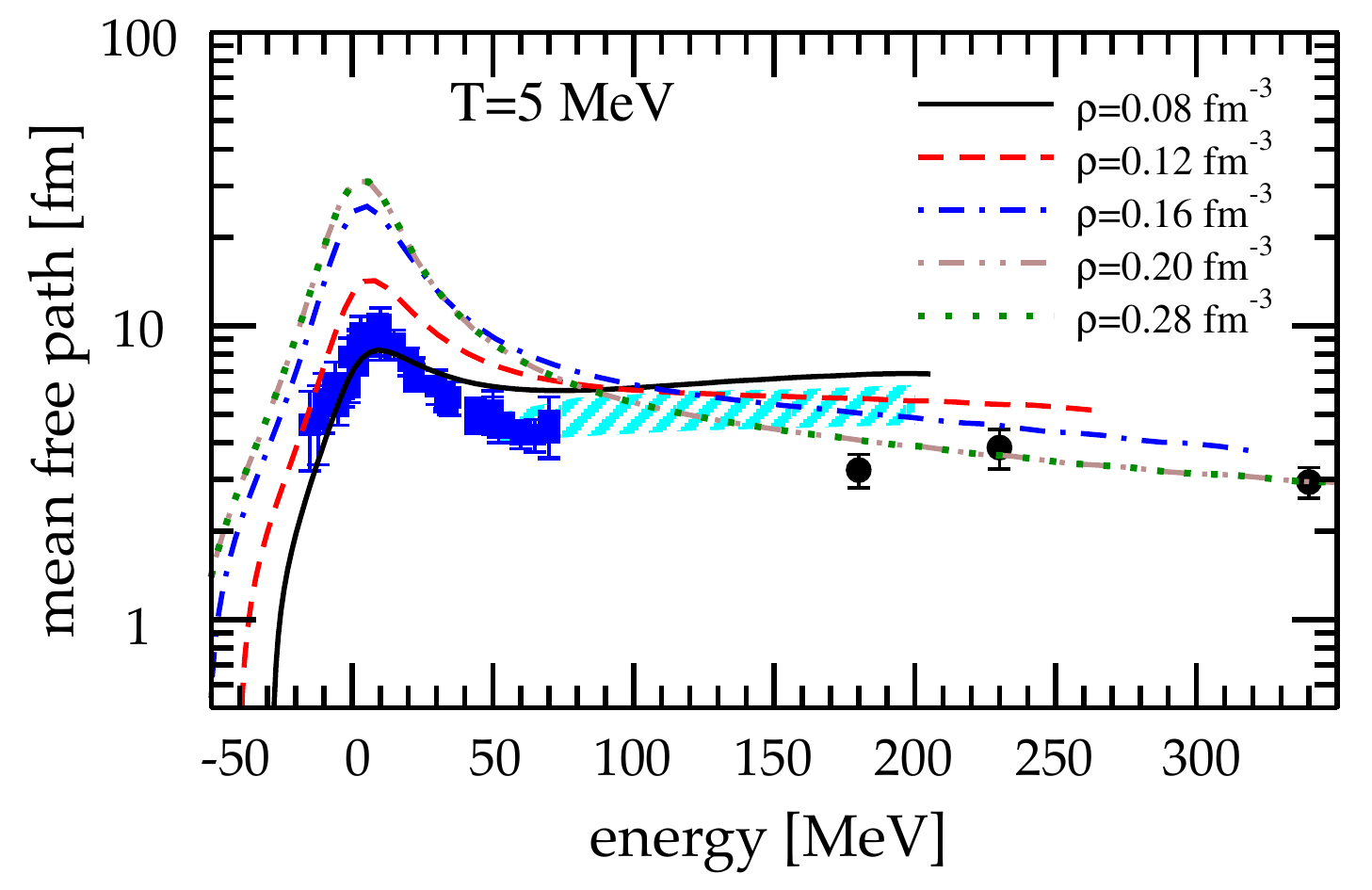}
\end{center}
\caption{Nucleon mean free path in (isospin symmetric) nuclear matter. Theoretical calculations performed within the self-consistent Green's function approach (lines)~\cite{Rios12} are compared to an evaluation of different experimental data (dots, squares and shaded area)~\cite{Lopez14}. Figure adapted from Ref.~\cite{Rios12}.}
\label{fig_mfp}
\end{figure}

If the constituents of a many-body system do not interact, the total Hamiltonian can be written as a sum of one-body terms,
\begin{equation}
H =  \sum_i^A h_i \; .
\label{eq_H_ip}
\end{equation}
The corresponding $A$-body problem then reduces to $A$ one-body problems,
\begin{equation}
H \ket{\Psi^A_k} = E^A_k  \ket{\Psi^A_k} 
\;\;\;\; \longrightarrow \;\;\;\;
h_i \ket{\phi^i_k} =  \varepsilon^i_k \ket{\phi^i_k}  \; ,
\label{eq_Sch_ip}
\end{equation}
where $\ket{\phi^i_k}$ denotes the one-body wave function of the $i$-th nucleon.
Clearly, this would greatly simplify our task both from the conceptual and from the practical point of view.
However, we know that nucleons are interacting and the strong force between them is at the origin of nuclear binding itself, i.e. at the very existence of such systems.
Moreover, the inter-nucleon distance in nuclei is roughly 2 femtometers (fm), about the same as the range of nuclear interactions. 
Does an independent-particle picture make sense at all?
Surprisingly, it turns out it does. 
First, Fermi statistics helps out. 
Second, the mean free path of nucleons in nuclear matter is rather large.
This is supported by recent measurements and theoretical calculations, see Fig.~\ref{fig_mfp}.

If nucleons can thus be considered, to a first approximation, as independent particles in a common (one-body) potential well that binds them to form the nucleus, the one-body Hamiltonians in Eqs.~\ref{eq_H_ip} and~\ref{eq_Sch_ip} can be written as
\begin{equation}
h_i = \frac{p_i^2}{2m} + V(r_i) \; ,
\label{eq_hi}
\end{equation}
where the first term on the right-hand side represents the nucleon kinetic energy and the second one the one-body potentials.
Commonly used forms are potentials of the Woods-Saxon type
\begin{equation}
V(r_i) = - \frac{V_0}{1+\exp{(\frac{r_i-R}{a})}} \; ,
\label{eq_WS}
\end{equation}
with $V_0$, $R$ and $a$ constants to be suitably adjusted.
\begin{figure}[t]
\begin{center}
\includegraphics[width=7cm]{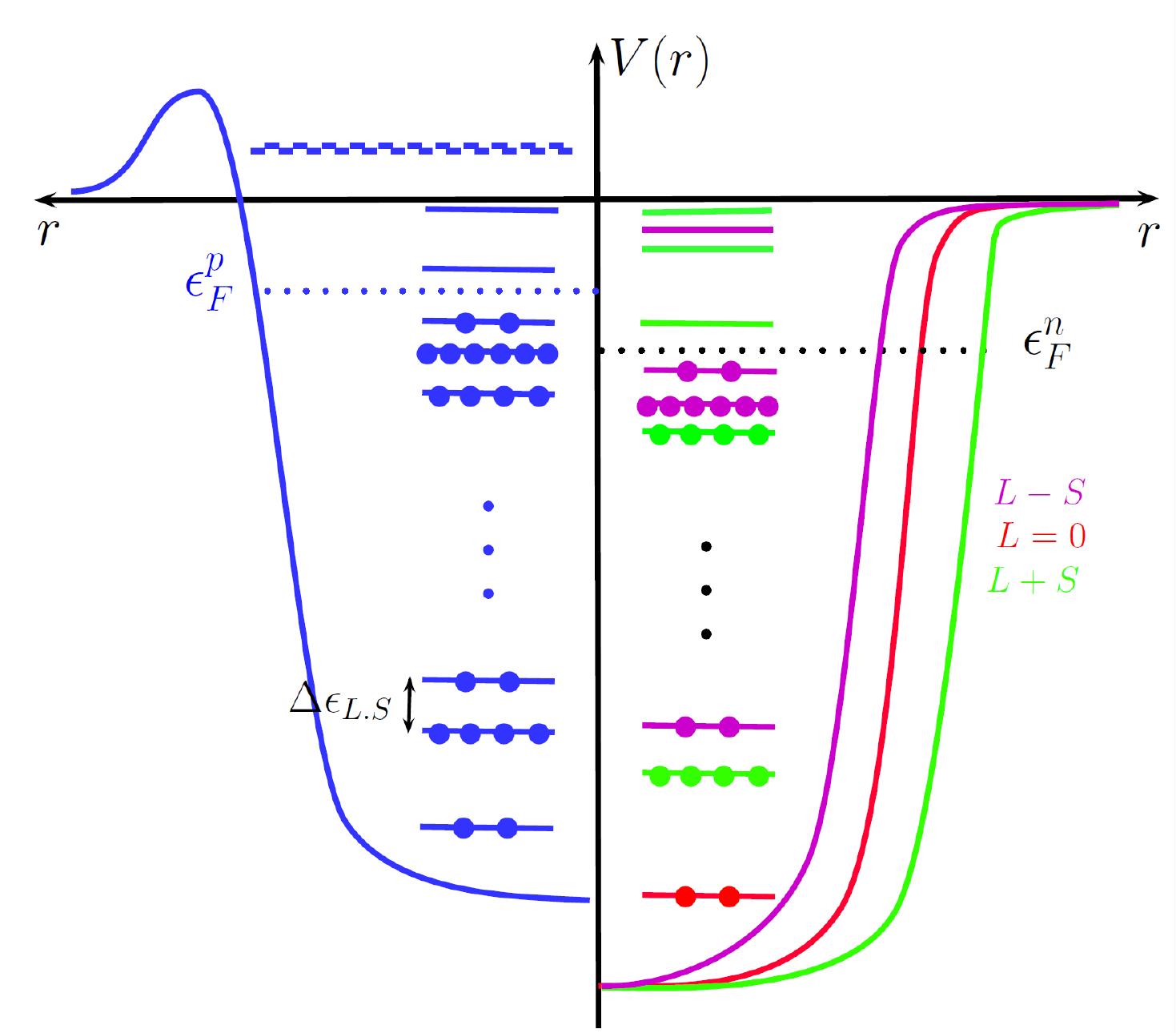}
\end{center}
\caption{Pictorial representation of nucleonic energy levels in the independent particle picture. Proton and neutron Fermi energies are denoted by $\epsilon_F^p$ and $\epsilon_F^n$ respectively. Notice that Coulomb interactions shift the proton potential up.}
\label{fig_well}
\end{figure}
Nucleons are then placed in successive energy levels (corresponding to the $\varepsilon^i_k$ in Eq.~\eqref{eq_Sch_ip}) according to Pauli principle, see a pictorial representation in Fig.~\ref{fig_well}. 

Let us remark that nuclei are self-bound systems. 
Hence, there is no external potential that keeps the nucleons confined like e.g. the Coulomb potential for electrons in an atom. 
This common potential instead originates as a sort of average from the individual interactions between protons and neutrons, and for this reason it can be denoted as \textit{mean field}.

\subsection{Shell model}
\label{sec_sm}

\subsubsection{Non-interacting shell model}

\begin{figure}[h]
\begin{center}
\includegraphics[width=8.5cm]{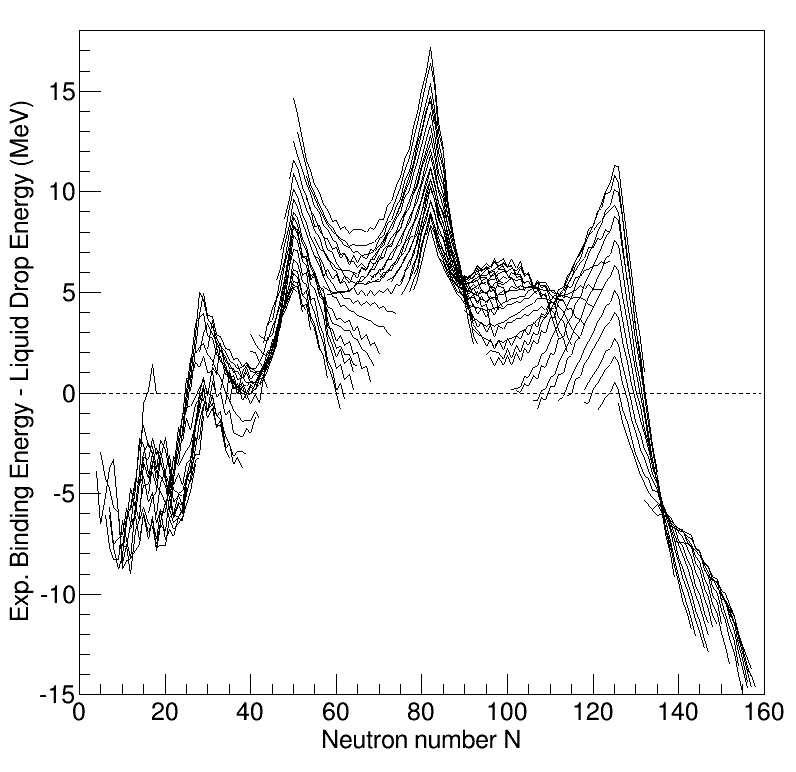}
\end{center}
\caption{Difference between binding energies computed with the semi-empirical mass formula~\eqref{eq_semi} and experimental values. Results are displayed for different isotopic chains as a function of neutron number. Taken from Ref.~\cite{wikiNS}.}
\label{fig_shells}
\end{figure}
How exactly do nucleons get organised in the underlying one-body potential?  Are energy levels distributed evenly, randomly? 
A hint comes from comparing predictions from the mass formula~\eqref{eq_semi} to measured binding energies, see Fig.~\ref{fig_shells}.
One notices that, for different isotopic chains, deviations follow certain patterns as a function of neutron numbers.
Specifically, the distributions are peaked at fixed values of $N$=8, 20, 28, 50, 82, 126. 
These correspond to particularly stable configurations of $N$ (and $Z$) that at first resulted completely unexplained, whence their denomination as \textit{magic} numbers.
What creates these regular patterns?

This picture recalls the one of electron shells in the atom, where jumps in the distribution of energy levels reflect the filling of the various shells.
Yet, as opposed to the atomic case governed by the (external) Coulomb potential, no obvious common potential can be identified for nucleons.
First attempts to design an effective one-body potential were based on a three-dimensional harmonic oscillator complemented with a term proportional to the angular momentum squared to account for centrifugal effects.
However they were not able to account for the observed magic numbers.
In 1949 Goeppert-Mayer and Jensen\footnote{For this work, Maria Goeppert-Mayer and Hans Jensen were awarded the 1963 Nobel prize.} proposed the inclusion of an additional term that describes the spin-orbit interaction~\cite{goeppertmayer49,Haxel49}.
This resulted crucial to reproduce the pattern of magic numbers and defined what is known as the (non-interacting) nuclear shell model.
\begin{figure}[h]
\begin{center}
\hspace{-0cm}
\includegraphics[width=6.5cm]{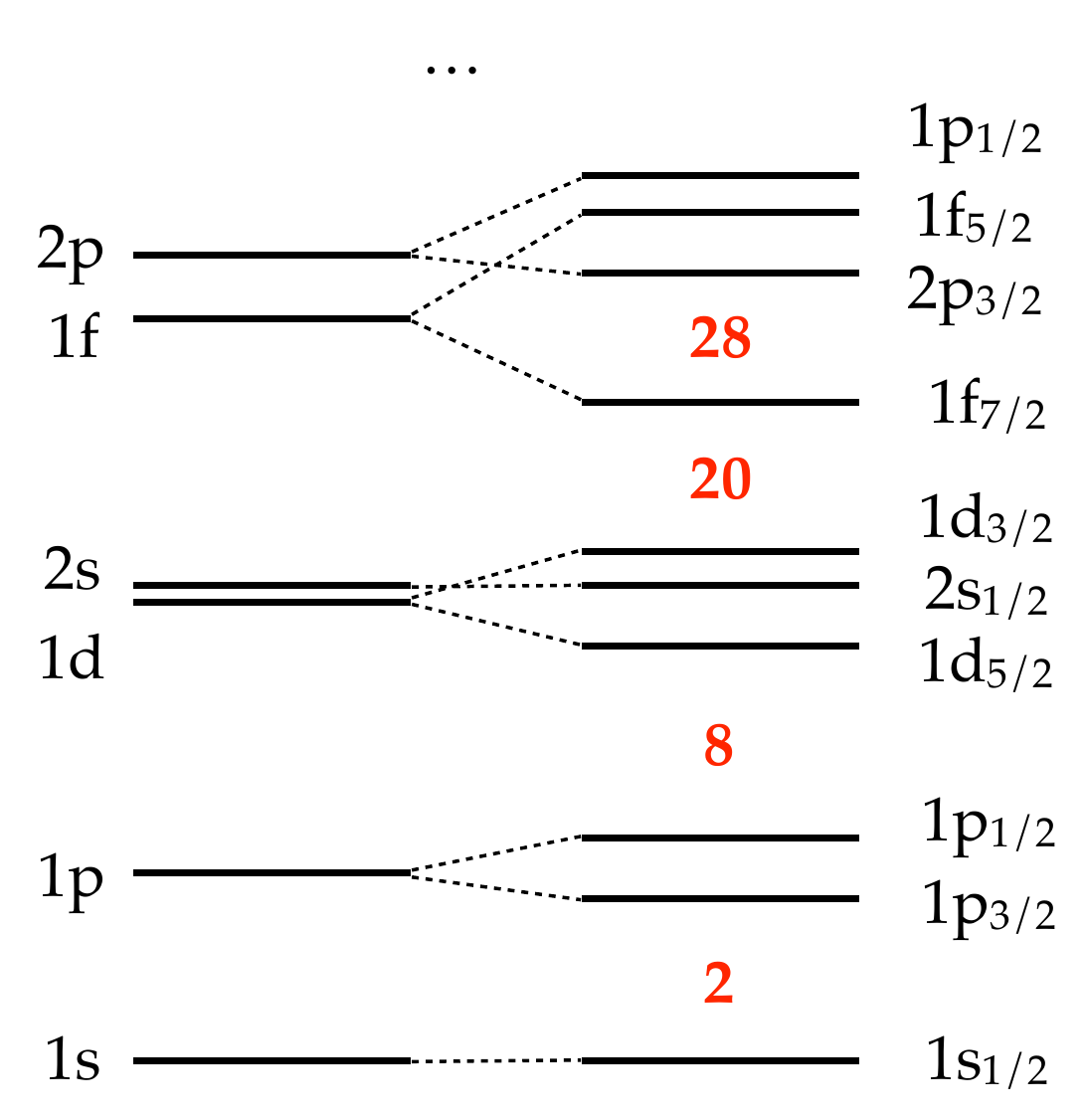}
\end{center}
\caption{Schematic organisation of harmonic-oscillator energy levels. On the right, it is depicted how the inclusion of a spin-orbit potential creates the splitting that allows to correctly describe stable magic numbers.}
\label{fig_ISM}
\end{figure}

Figure~\ref{fig_ISM} illustrates the organisation of single-particle levels and the appearance of magic numbers.
Nucleons\footnote{Two separate sets of levels host protons and neutrons.} are placed in levels of increasing energy, starting with the one corresponding to the lowest possible energy. 
Such energies depend on the principal quantum number $n$, the orbital angular momentum $\ell$ and, once the spin-orbit term is added, on the total angular momentum $j$.
However, because of the rotational symmetry of the Hamiltonian, they are independent of the angular momentum projection $m_j$.
This generates a degeneracy of $2j+1$ for each level with momentum $j$, e.g. the $1\text{p}_{3/2}$ level\footnote{The spectroscopic notation $n \ell_j$ is usually employed to label each energy level, with $\ell = \{ \text{s,p,d, ...} \}$ standing for $\ell = \{ 0,1,2, ... \}$.} can host four neutrons (or protons, depending which set we are considering), and so on.
By looking at the level schemes of Fig.~\ref{fig_ISM}, one sees that large energy gaps are present between certain groups of levels. 
Such groups are called (major) \textit{shells}. 
When a shell is full, i.e. all corresponding levels are filled with nucleons, one talks about a \textit{closed} shell\footnote{Level schemes where both neutron and protons form closed shells are called \textit{doubly closed shells}.}.
Because of the large energy gap, closed-shell nuclei result in particularly stable configurations. 
Consequently, the corresponding numbers of neutrons or protons are denoted as \textit{magic numbers}.
On the right hand side of Fig.~\ref{fig_ISM}, one sees that split induced by the spin-orbit potential rearranges the major shells in such way that the experimentally observed stable configurations match the magic numbers resulting from the level scheme.

The concepts of mean field and nucleon shells played a pivotal role in the subsequent development of nuclear models and shaped our understanding of the atomic nucleus.
In fact, reference to an underlying single-particle spectrum is often made to explain the characteristics and the evolution of low-energy observables in nuclei.
However, one must keep in mind that nuclei are quantum many-body systems in which individual nucleons do \textit{not} occupy stationary single-particle states. 
In other words, the physical observables are the ones obtained by solving the \textit{many-body} Schr{\"o}dinger equation (left-hand side of Eq.~\eqref{eq_Sch_ip}), not an auxiliary \textit{one-body} problem (right-hand side of Eq.~\eqref{eq_Sch_ip}).
Nucleon shells are thus a useful interpretative tool but \textit{non observable} in the quantum mechanical sense. 
This implies that single-nucleon energy levels (usually referred to as effective single-particle energies) are model dependent and can not be compared to measured experimental data in a meaningful way.
Recently, the non observable nature of the nuclear shell structure was discussed and illustrated in the framework of ab initio calculations~\cite{Duguet15}.

\subsubsection{Interacting shell model}

While an independent-particle model is sufficient to explain basic features of nuclear structure like magic numbers and the existence of closed-shell nuclei, in general a correlated wave function is needed. 
That is, one can not have an exact correspondence of the type depicted in Eq.~\eqref{eq_Sch_ip} but instead decompose the total Hamiltonian in a mean-field and a residual part, 
\begin{equation}
H = H_{\text{MF}} + H_{\text{res}}  \; .
\end{equation}
Whereas $H_{\text{MF}}$ can be easily handled and yields an independent-particle picture of the type depicted in Fig.~\ref{fig_ISM}, $H_{\text{res}}$ requires a more sophisticated treatment, e.g. a diagonalisation in the $A$-body Hilbert space.
This second part may become extremely costly as $A$ increases.

\begin{figure}[t]
\begin{center}
\hspace{-0cm}
\includegraphics[width=5.5cm]{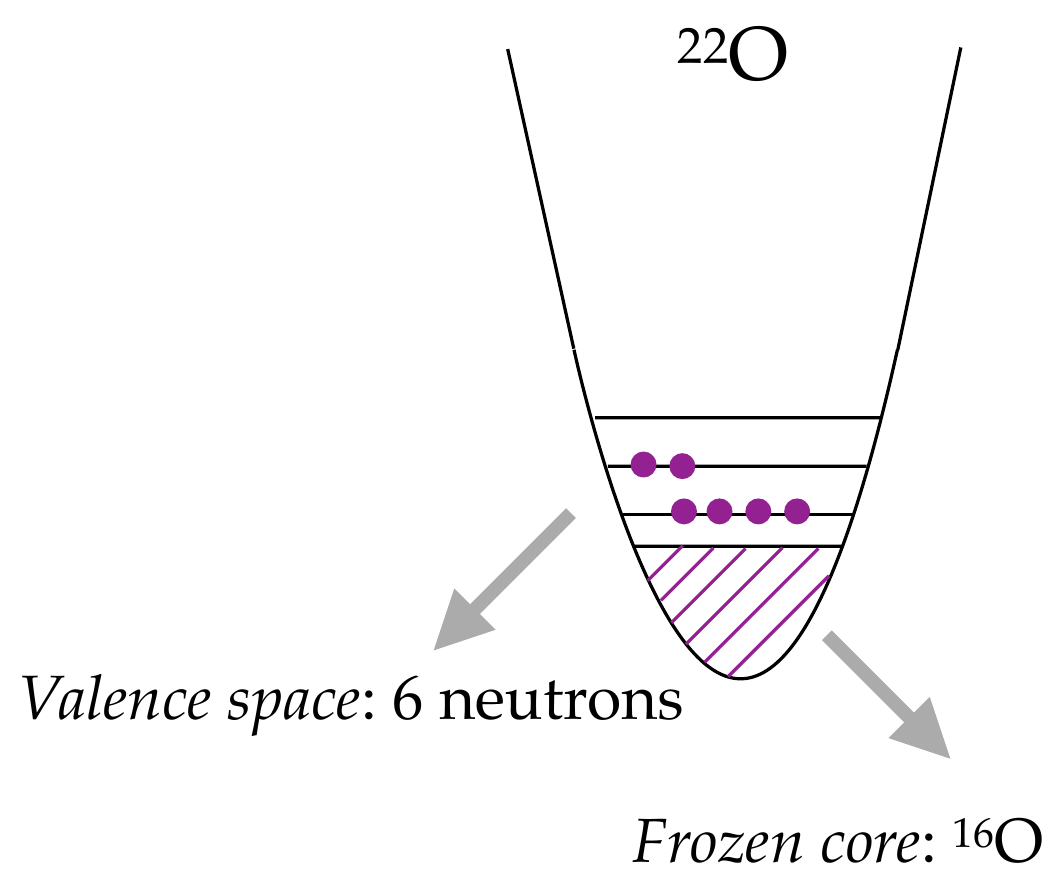}
\end{center}
\caption{Schematic view of the interacting shell applied to $^{22}$O.}
\label{fig_SM}
\end{figure}

In order to overcome this ``curse of dimensionality'' one might think of exploiting the large energy separation between major shells.
A large energy separation implies that excitations across major shells are suppressed.
Then, to a first approximation, a given major shell can be investigated independently of the others.
This idea is at the basis of the interacting shell model and is implemented in three steps:
\begin{enumerate}
\item Define an active Hilbert space (called the \textit{valence} space). Typically, it includes a full major shell (i.e. well separated in energy from other shells).
\item Build a valence-space Hamiltonian $H^{\text{eff}}_{\text{val}}$.
\item Diagonalise $H^{\text{eff}}_{\text{val}}$ in the reduced space.
\end{enumerate}
In doing so, one can address systems that would be out of reach for a diagonalisation of the full Hilbert space.
An example for one oxygen isotope, $^{22}$O, is depicted in Fig.~\ref{fig_SM}.

\begin{figure}[h]
\begin{center}
\hspace{-0cm}
\includegraphics[width=7cm]{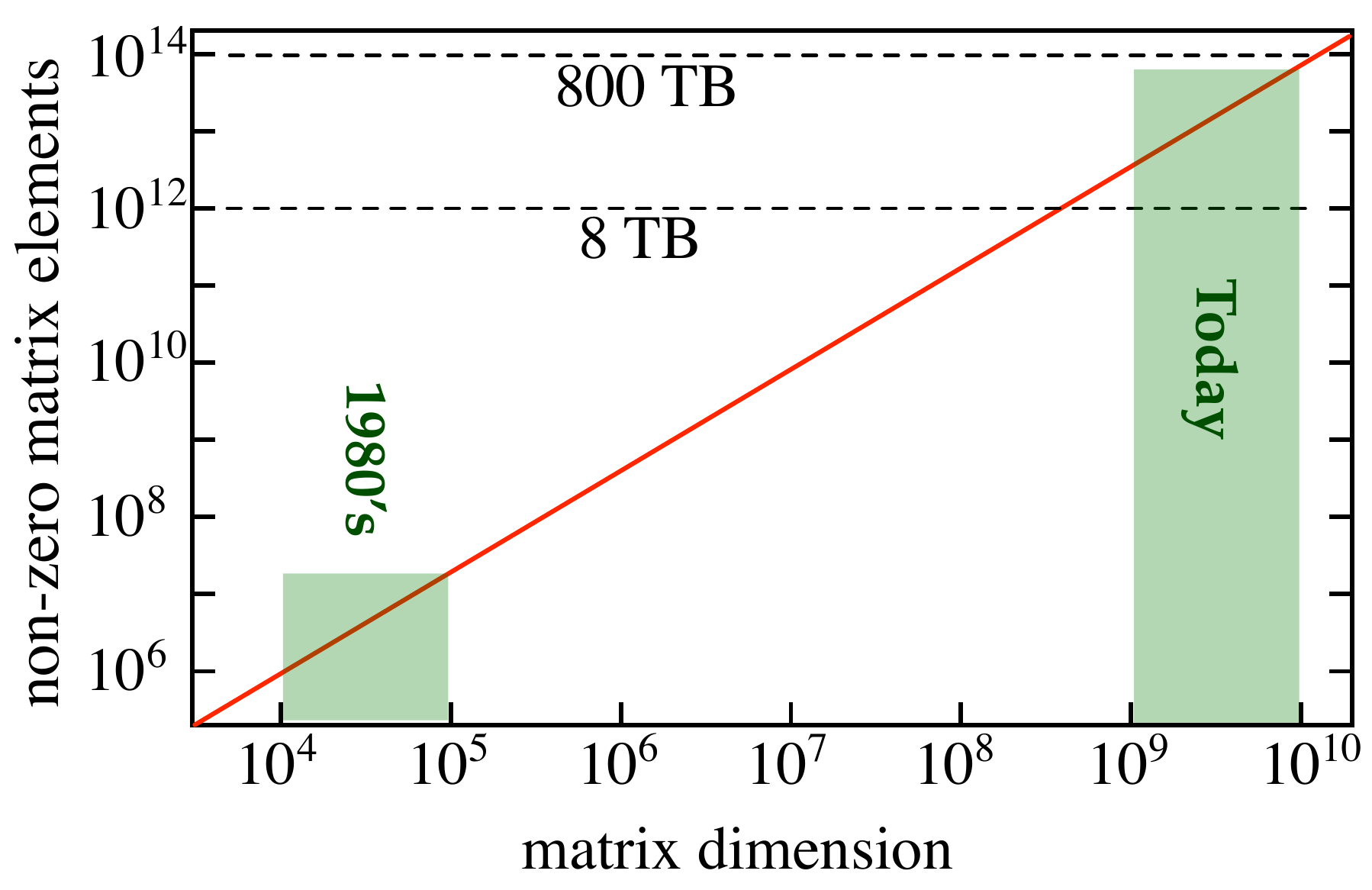}
\end{center}
\caption{Approximate dependence of the number of non-zero matrix elements on the matrix dimension for shell-model Hamiltonians. Examples of corresponding needs in aggregate memory are shown, together with the values reachable in 1980s and with current state-of-the-art high-performance computational resources (green bands).}
\label{fig_diag_dim}
\end{figure}
How is $H^{\text{eff}}_{\text{val}}$ constructed? 
There exist two alternative ways.
The first one consists in fitting the parameters of $H^{\text{eff}}_{\text{val}}$ to available experimental data.
In doing so, excellent accuracy can be reached locally for refined spectroscopy.
The drawbacks are that different Hamiltonians have to be used for different valence spaces and that the quality of the description deteriorates when approaching regions of the Segr\`e chart in which experimental information is scarse or not available.
The second possibility consists in deriving it from the original Hamiltonian via some Hilbert-space projection techniques. 
In doing so, the link with ``microscopic'' inter-nucleon forces is preserved and the approach is systematic and universal.
The drawback is that such a projection requires sophisticated many-body techniques that might themselves lead to unfavourable scaling and costly calculations.

The first route, the phenomenological shell model, has been developed and applied for over 50 years, and is today one of the reference methods in nuclear structure~\cite{Caurier05}.
Implementations of the second option, the so-called valence-space ab initio or ab initio shell-model approach, have appeared only very recently~\cite{Bogner14,Jansen14} as a byproduct of the recent progress in ab initio techniques (see Section~\ref{sec_abinitio}).

As discussed, the interacting shell model allows performing calculations well beyond the region reachable by full-space diagonalistion. 
Nevertheless, while the concept of well-separated shells is valid in light and medium-mass nuclei, its utility becomes limited in heavier systems. 
There, shells are either close to one another or contain a large number of nucleons, such that even a valence-space diagonalistion becomes at some point prohibitive, see Fig.~\ref{fig_diag_dim}.
As of today shell model applications are limited to $A \sim 100$.

\subsection{Energy density functionals}
\label{sec_EDF}
The other possibility of simplifying the full Schr{\"o}dinger equation consists in making a simple ansatz for the many-body wave function. 
As a result\footnote{The aim is of course to have the same eigenvalues of the full Schr{\"o}dinger equation in the end.}, many-body correlations must be somehow incorporated into the Hamiltonian $H^{\text{eff}}$.
The first attempts in this direction used the full original Hamiltonian\footnote{Let us remark that there is no ``full original Hamiltonian'' in absolute, but only possible models of it. Hence here we refer to the models available at that time.} as a starting point and Hartree-Fock (HF) theory to generate a mean-field potential in a self-consistent way. 
The approach resulted not very satisfactory, both because of the poor quality of the available Hamiltonians and, mainly, because the HF approximation was too simple to be able to capture all the necessary correlations.

Roughly speaking, HF theory can be seen as a convolution between a two-body interaction and a one-body density matrix, which generates a density-dependent one-body (i.e. mean-field) potential. 
A second convolution with the density matrix yields the interaction energy.
Following the first Hamiltonian-based attempts, people naturally started adjusting the different parameters of the starting two-body interaction in order to match experimental data.
This led to designing a phenomenological Hamiltonian, whose parameters effectively account for many-body correlations. 
The following step was to abandon the link with an underlying Hamiltonian and directly postulate the total energy of the system as a general functional of the (one-body) density, whence the name \textit{energy density functional} (EDF).

The general character of the functional provides a good deal of flexibility, which is necessary to account for the complicated inter-nucleon correlations. 
In addition, what makes the EDF approach very powerful is the fact that it exploits the concept of symmetry breaking and restoration.
The idea is the following.
In consequence of Schr{\"o}dinger equation, the (exact) many-body wave function conserves certain symmetries.
Symmetry-\textit{conserving} approaches retain such symmetries while designing approximations to the exact wave function and computing the corresponding energies. 
Symmetry-\textit{breaking} approaches lift this requirement and allow one or more symmetry to be broken in designing the approximate wave function. 
Usually, at a given level of approximation, the energies resulting from the latter are closer to the exact ones than the former.
Since physical solutions do have the symmetries of the Hamiltonian, broken symmetries have to be eventually restored at the end of the calculation.
This second step is typically achieved by mixing various symmetry-broken solutions and projecting out the component corresponding to the good symmetry.
These two steps of the EDF method are called respectively single-reference (SR) and multi-reference (MR)\footnote{The EDF approach developed in nuclear physics has some common points with the density functional theory (DFT) formalised by Kohn, Hohenberg and Sham and widely used in quantum chemistry. However, EDF differs from DFT precisely because of this essential recourse to symmetry breaking and restoration.}.

Several different EDF implementations have been developed over the years. 
The two main non-relativistic approaches, the zero-range Skyrme and the finite-range Gogny functionals were conceived in the 1970s and are still actively developed and used in state-of-the-art applications~\cite{bender03b}.
Relativistic approaches date back to the 1980s.
In all of them, the strategy is to fit the parameters (about 15 in modern functionals) at the SR level to a given set of experimental data, usually a selection of binding energies and charge radii.
From the computational point of view, the SR step is inexpensive and scales very gently with the mass number, such that systematic SR calculations across the Segr\`e chart are at hand for any EDF implementation. 
In fact, at present SR-EDF constitute the \textit{only} practicable microscopic method able to address the whole nuclear chart and thus predict how many nuclei actually exist, as exemplified in Fig.~\ref{fig_segre}.
MR calculations exist in different variants and are generally much more costly. 
As a result, systematic surveys of the whole chart remain very challenging and state-of-the-art MR approaches are typically limited to selected cases of high experimental interest~\cite{Bally14,Egido16}.
Nevertheless, it is the only microscopic approach available to study super- and hyper-heavy nuclei.

The main drawback of such methods resides in their non systematic character. 
On the one hand, different functionals might perform equally well when compared to existing experimental data but yield diverging predictions where data is unavailable. 
On the other hand, it is not always obvious how one to improve their design. 
Recently it was also realised that some problematic pathologies are associated to MR calculations, where the ad hoc use of nuclear densities lacks rigorous roots~\cite{Lacroix09,Bender09,Duguet09}. 
These pathologies are not present if one follows the original route that requires starting with a Hamiltonian operator. 
In order to tackle these shortcomings, novel formal developments are being proposed on several aspects including possible connections with ab initio approaches~\cite{Duguet15b,Dobaczewski16}.

%%%%%%%%%%%%%%%%%%%%%
\section{Early models of NN interactions}
\label{sec_old}
%%%%%%%%%%%%%%%%%%%%%

\subsection{Basic properties of NN interactions}
\label{sec_properties}

The modelling of the nuclear Hamiltonian has kept nuclear theorists occupied for over 80 years and, although significant advances have been made, it has not been settled yet.
Before discussing early and modern Hamiltonian models, let us briefly examine which basic properties it should have and how we can get specific information about it.
The simplest and most natural thing is to start by considering the two-nucleon system, where a general Hamiltonian reads
\begin{equation}
H=T+V_{NN}+V_{\text{em}} \; .
\label{eq_v2}
\end{equation}
Here $T$ represents the kinetic energy operator, $V_{NN}$ the nuclear two-body interaction and $V_{\text{em}}$ the electromagnetic potential (i.e. Coulomb plus small corrections). 
The latter being known and the first being trivial, we are interested in constructing the second term, which incorporates the strong interaction between two nucleons.

Relevant quantum numbers to describe a nucleon are, in addition to its position $\cc{r}$ and its momentum $\cc{p}$, its spin $\cc{$\sigma$}$ and its isospin $\cc{$\tau$}$.
Hence the most general form of the two-nucleon potential is
\begin{equation}
V_{NN} = V(\cc{r}_1,\cc{r}_2,\cc{p}_1,\cc{p}_2,\cc{$\sigma$}_1,
\cc{$\sigma$}_2,\cc{$\tau$}_1,\cc{$\tau$}_2) \; .
\label{eq_vqn}
\end{equation}
Several symmetries can then be exploited to constrain this form, either continuous (translation in time/space, rotation in space/spin, Galilean invariance) or discrete (parity, time reversal, baryon + lepton-number conservation).
Still remaining general, we can write
\begin{equation}
V_{NN} = V_1(\cc{r},\cc{p},\cc{$\sigma$}_1, \cc{$\sigma$}_2)
+ V_\tau(\cc{r},\cc{p},\cc{$\sigma$}_1, \cc{$\sigma$}_2) \, \cc{$\tau$}_1 \cdot \cc{$\tau$}_2 \; ,
\label{eq_vtra}
\end{equation}
where
\begin{equation}
\begin{aligned}
\cc{r} &= \cc{r}_1-\cc{r}_2 \; , \\
\cc{p} &= \cc{p}_1-\cc{p}_2 \; ,
\label{eq_rp}
\end{aligned}
\end{equation}
and each of the two terms contains a spin-scalar, a spin-vector and a spin-tensor part.
Furthermore, to a first approximation the potential can be considered\footnote{This type of information can be deduced e.g. by studying properties of mirror nuclei, i.e. nuclei with $(Z,N) = \{ (X,Y), (Y,X) \}$, and proton-proton vs proton-neutron scattering.} \textit{charge symmetric}, i.e. $V_{pp} \approx V_{nn}$, and \textit{charge independent}, i.e. $V_{pp} \approx V_{pn} \approx V_{nn}$.

\begin{figure}[t]
\begin{center}
\hspace{-0cm}
\includegraphics[width=8.5cm]{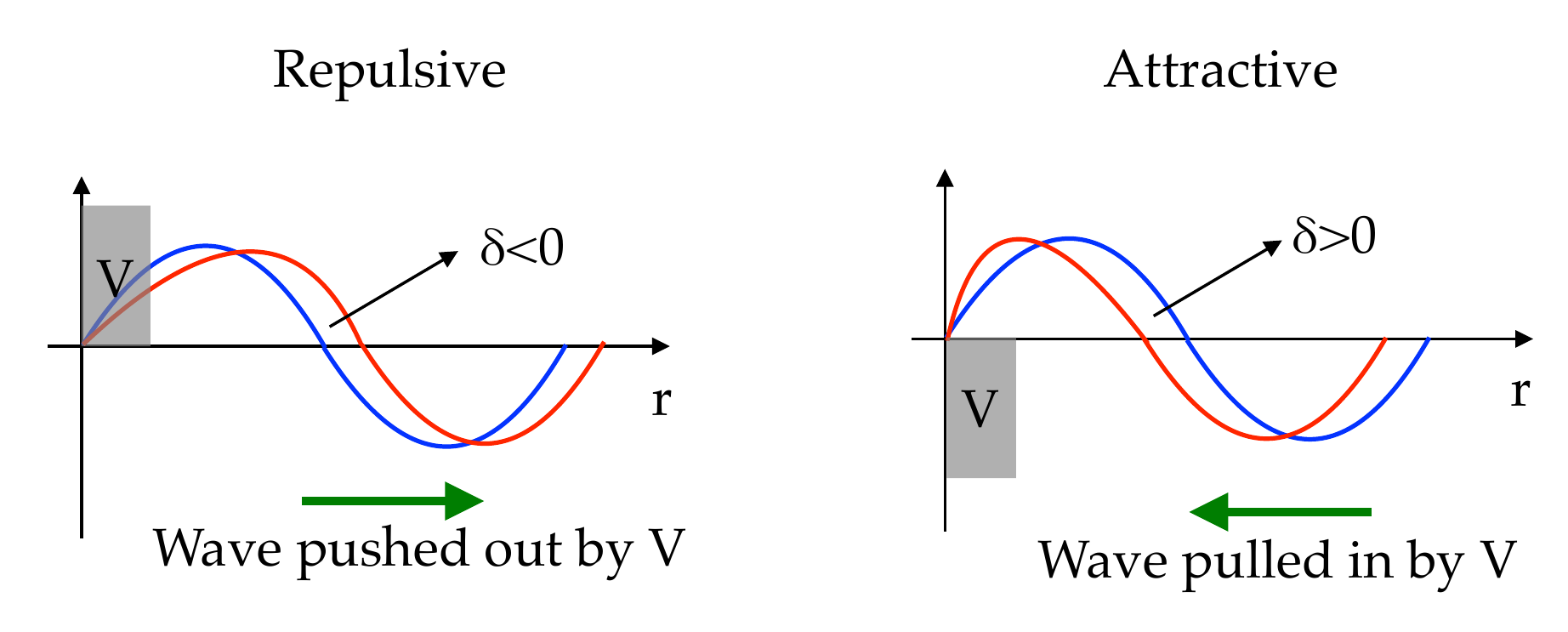}
\end{center}
\caption{Illustration of scattering phase shifts. In the case of a repulsive interaction (left), the scattered wave (red) is pushed out by the potential $V$, which generates a shift $\delta< 0$ in its phase. The converse happens for an attractive potential (right).}
\label{fig_phaseshifts}
\end{figure}

From these simple considerations one can already see that $V_{NN}$ will be constituted by several terms. 
How can we gain information on them?
The first source of information is nucleon-nucleon scattering.
In particular, when two particles collide, the interaction leads to a change in the phase of the scattered wave. 
By examining these \textit{scattering phase shifts} many details about the interaction between the two particles can be inferred, starting from its attractive or repulsive nature, see Fig.~\ref{fig_phaseshifts}.

\begin{figure}[t]
\begin{center}
\hspace{-0cm}
\includegraphics[width=8cm]{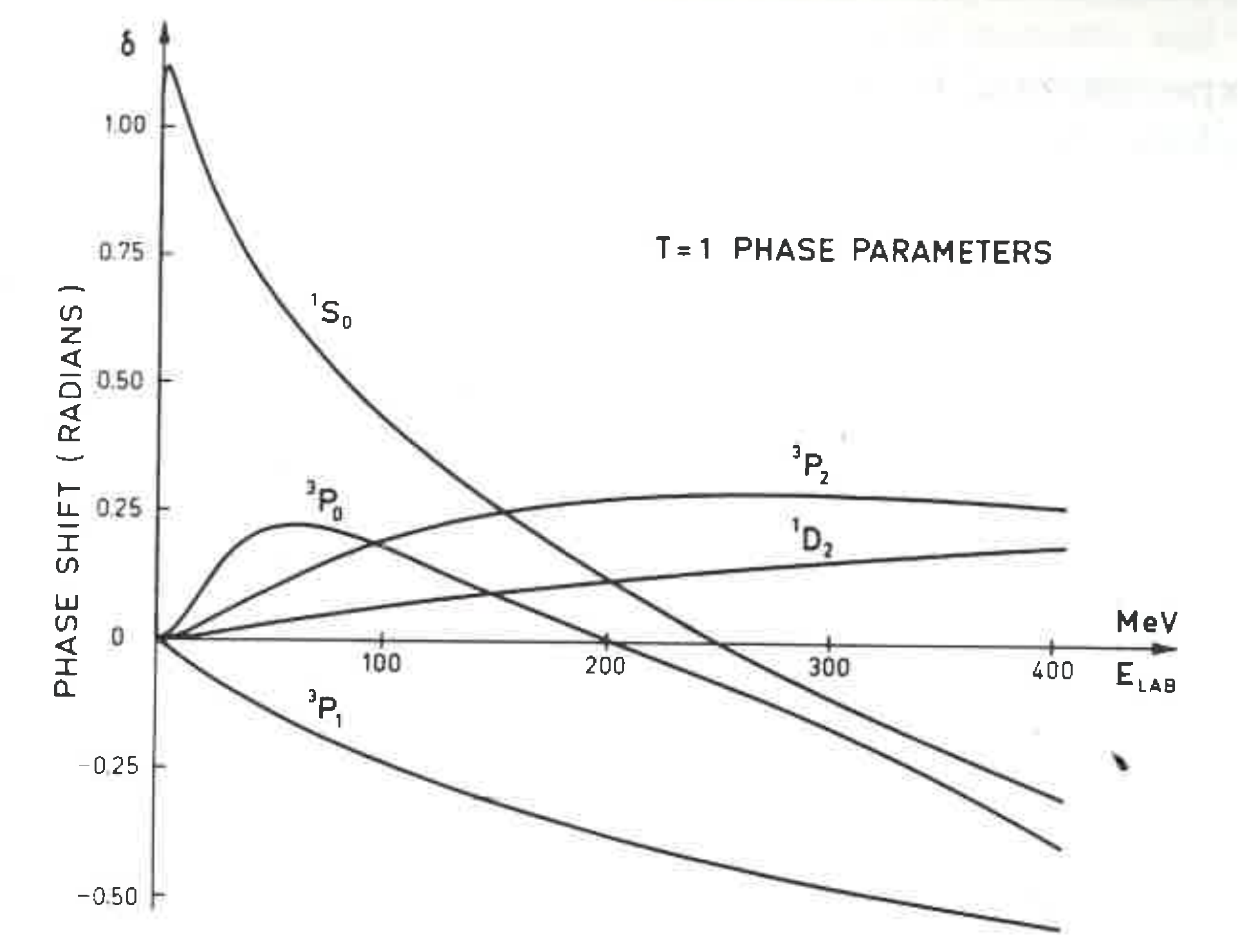}
\end{center}
\caption{Experimental proton-neutron phase shifts in different partial wave channels as a function of the scattering energy.}
\label{fig_ps_example}
\end{figure}

Scattering phase shifts are conveniently studied in a partial-wave\footnote{The partial wave notation is $^{2S+1}L_J$, where $J$, $L$ and $S$ are respectively the total angular momentum, the orbital angular momentum and the spin of the pair of scattering nucleons.}
%\footnote{\textcolor{blue}{Introduce partial waves. 
%\begin{equation}
%\vec{J} = \vec{L} + \vec{S} \,\, \Longrightarrow \,\, |L-S| \leq J \leq |L+S| 
%\label{eq_pw1}
%\end{equation}
%\begin{equation}
%\vec{S} = \vec{s_1} + \vec{s_2} \,\, \Longrightarrow \,\, S= 0,1
%\label{eq_pw2}
%\end{equation}
%\begin{equation}
%J =
%\left\{
%\begin{array}{rl}
% L & \mbox{for } S=0 \\%
%|L-1|,L,L+1 & \mbox{for } S=1
%\end{array}
%\right.
%\label{eq_pw3}
%\end{equation}
%}} 
%\end{comment}
analysis.
An example is shown in Fig.~\ref{fig_ps_example}, where neutron-neutron phase shifts for different partial waves are displayed as a function of the collision energy in the laboratory frame. 
For instance 
\begin{itemize}
\item[$\circ$] From the fact that the $^1S_0$ phase shift is large at small energy\footnote{Although it might not be visible from the figure, it actually goes to zero at $E=0$.} one can deduce that there is nearly a bound state in this channel\footnote{The number of bound states in a given partial wave is related to the value of the corresponding phase shift at zero energy~\cite{Levinson49}.};\\
\item[$\circ$] From the fact that the $^1S_0$ phase shift becomes negative at high energies it follows that the potential, initially attractive, becomes repulsive at small distances;\\
\item[$\circ$] From the fact that $P$ waves are rather different from one another one deduces that there must be something else other than central terms (e.g. spin-orbit terms).
\end{itemize}

Another important source of information is the deuteron, i.e. the bound state of a proton and a neutron.
For instance, the fact that it has a non-zero quadrupole moment implies that the interaction must contain tensor terms.
By combining all these inputs, which have become richer and more precise over the years, nuclear physicists have been able to propose and construct more and more refined models of NN interactions.

\subsection{Yukawa potential}

The first model of nuclear forces dates back to the 1930s. 
At the time, the archetypical Coulomb interaction between charged particles was well known.
From the fact that its range is infinite, it follows that the carrier of the electromagnetic force, the photon, has zero mass.
From scattering experiments, however, the range of nuclear interaction appeared to be finite and was estimated to be about 2 fm.
The Japanese physicist Hideki Yukawa then had the intuition of replacing the zero-mass photon with a massive field of mass $m$ mediating interactions between nucleons (see Fig.~\ref{eq_yukawa}), and proposed what is nowadays known as Yukawa potential
\begin{equation}
V(r) \propto \frac{e^{-mr}}{r} \; .
\label{eq_yukawa}
\end{equation}
Assuming that the range of the potential is roughly equal to the Compton wavelength of the exchanged boson, i.e.
\begin{equation}
2 \text{ fm} \sim \frac{1}{m} \; ,
\label{eq_yukawa}
\end{equation}
Yukawa could estimate $m \sim 0.5 \text{ fm}^{-1} \sim100$ MeV.
\begin{figure}[b]
\begin{center}
\hspace{-0cm}
\includegraphics[width=5cm]{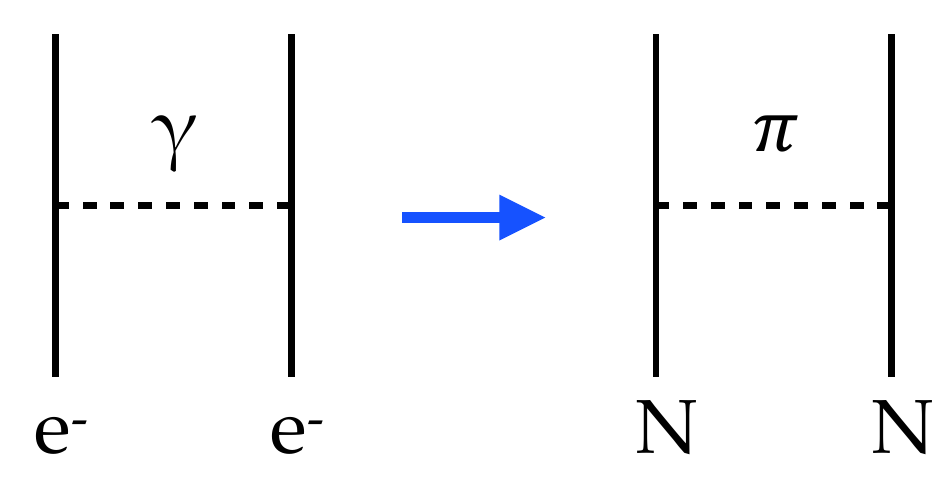}
\end{center}
\caption{Schematic view of Coulomb (left) and Yukawa (right) interactions.}
\label{fig_yukawa}
\end{figure}
Predicted in 1935 and originally named \textit{mesotron}, this boson was then identified with the pion, discovered in 1947 with a mass $m_\pi \approx 140$ MeV, not too different from Yukawa's value.

This idea proved to be extremely fitting and is still exploited in modern potentials to describe the \textit{long-range} part of the nuclear interaction.
Nevertheless, it quickly became evident that it could not be the full story and that the \textit{short-range} part had to be modelled in some other way. 
In the 1950s contributions from multi-pion exchange were studied with unsatisfactory results.
In the 1960s mesons heavier than the pion were discovered and, eventually, replaced multiple pion exchange in modelling the short-range part. 
This led to the development of various \textit{one-boson-exchange} (OBE) interactions.

\subsection{One-boson-exchange potentials}
Starting from the 1960s, heavier mesons such as $\rho, \omega$ and $\sigma$ began to be systematically included in the construction of NN interactions to model ranges smaller than $1/m_\pi$. 
Each of them generates a different spin/isospin structure that enriches the nuclear potential. 
Other terms compatible with the general symmetries of the Hamiltonian were then typically added \textit{ad hoc} to increase the flexibility of the model, in particular for what concerns the description of the short-range region.
Some others, notably the family of Argonne potentials, were entirely modelled as a sum of operators allowed by symmetries and general considerations.
\begin{figure}[b]
\begin{center}
\hspace{-0cm}
\includegraphics[width=7.5cm]{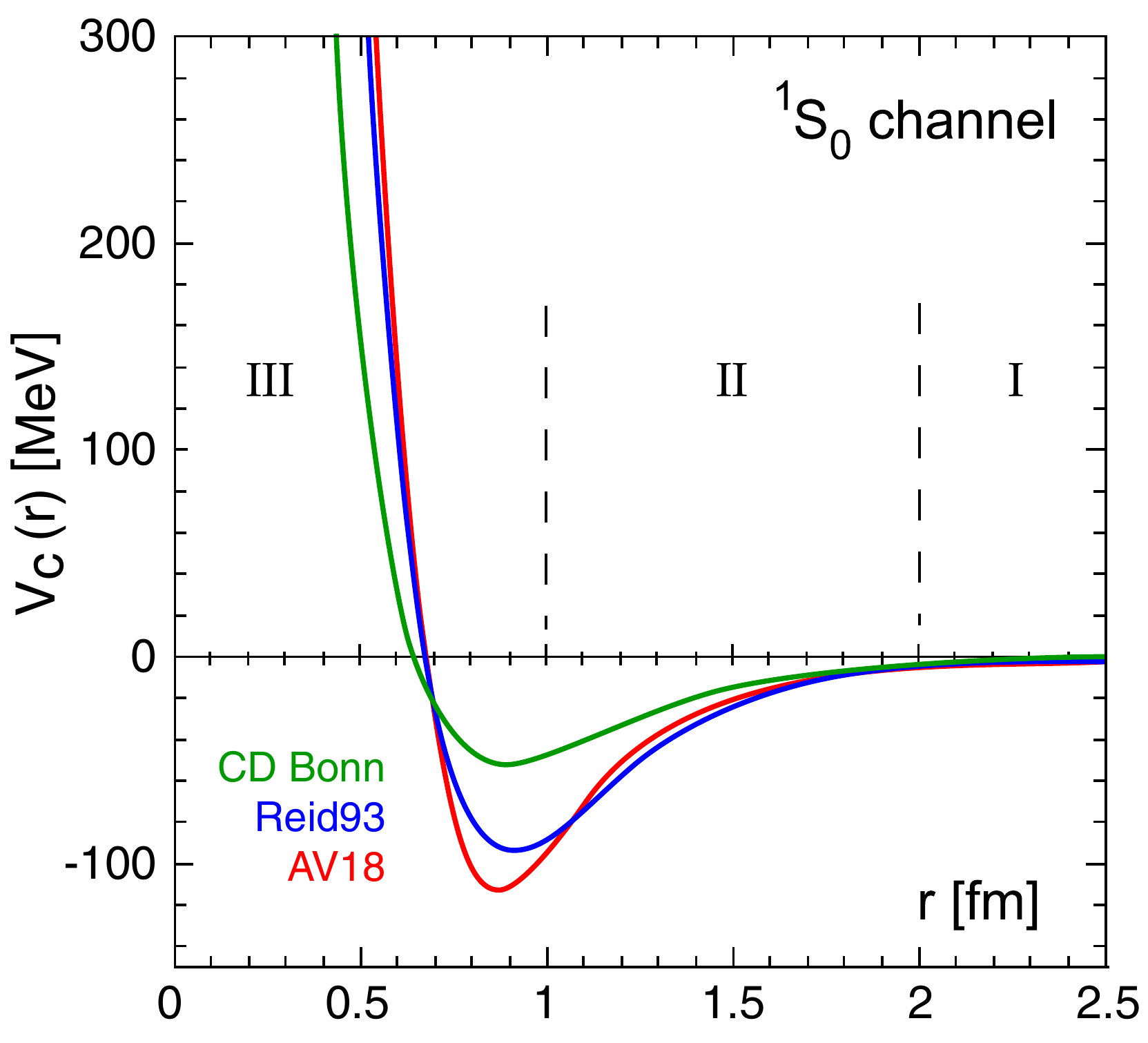}
\end{center}
\caption{Radial dependence of the central part of the nucleon-nucleon potential in the $^1$S$_0$ channel for three different Hamiltonian models. Regions corresponding to long-range (I), intermediate-range (II) and short-range (III) operators are roughly indicated.
Adapted from Ref.~\cite{Ishii07a}.}
\label{fig_obe}
\end{figure}

The common strategy of such models is the following:
\begin{enumerate}
\item Construct the operatorial structure of $V_{NN}$, i.e. spin/isospin scalar, vector or tensor terms and the respective radial functions;\\
\item Fit the coupling constants (usually appearing in the radial functions or as normalisation of individual terms) to experimental data, i.e. typically nucleon-nucleon scattering data and deuteron properties.
\end{enumerate}

The 1970s witnessed major developments in this context with several potentials appearing on the market.
Examples are the Paris, Bonn, Nijmegen and Argonne potentials.
Some of them are displayed, for one partial wave ($^1$S$_0$) and one operatorial structure (central), in Fig.~\ref{fig_obe}.
One notices that the various models differ in their radial behaviour, in particular in the short-range part.
This reflects the freedom to model high-energy details when interested in low-energy observables, an observation that is at the basis of effective field theories discussed in Section~\ref{sec_new}. 

Combining precise information becoming available on the experimental side and progress on the theoretical side, more and more refined potentials were developed up to achieving, for two-nucleon systems, a description with $\chi^2 / \text{datum} \approx 2$ in the 1980s and $\chi^2 / \text{datum} \approx 1$ in the 1990s.
Then, what about nuclear structure calculations?

\subsection{Three-nucleon forces}
\label{sec_3bf}

\begin{figure}[b]
\begin{center}
\hspace{-0cm}
\includegraphics[width=7.5cm]{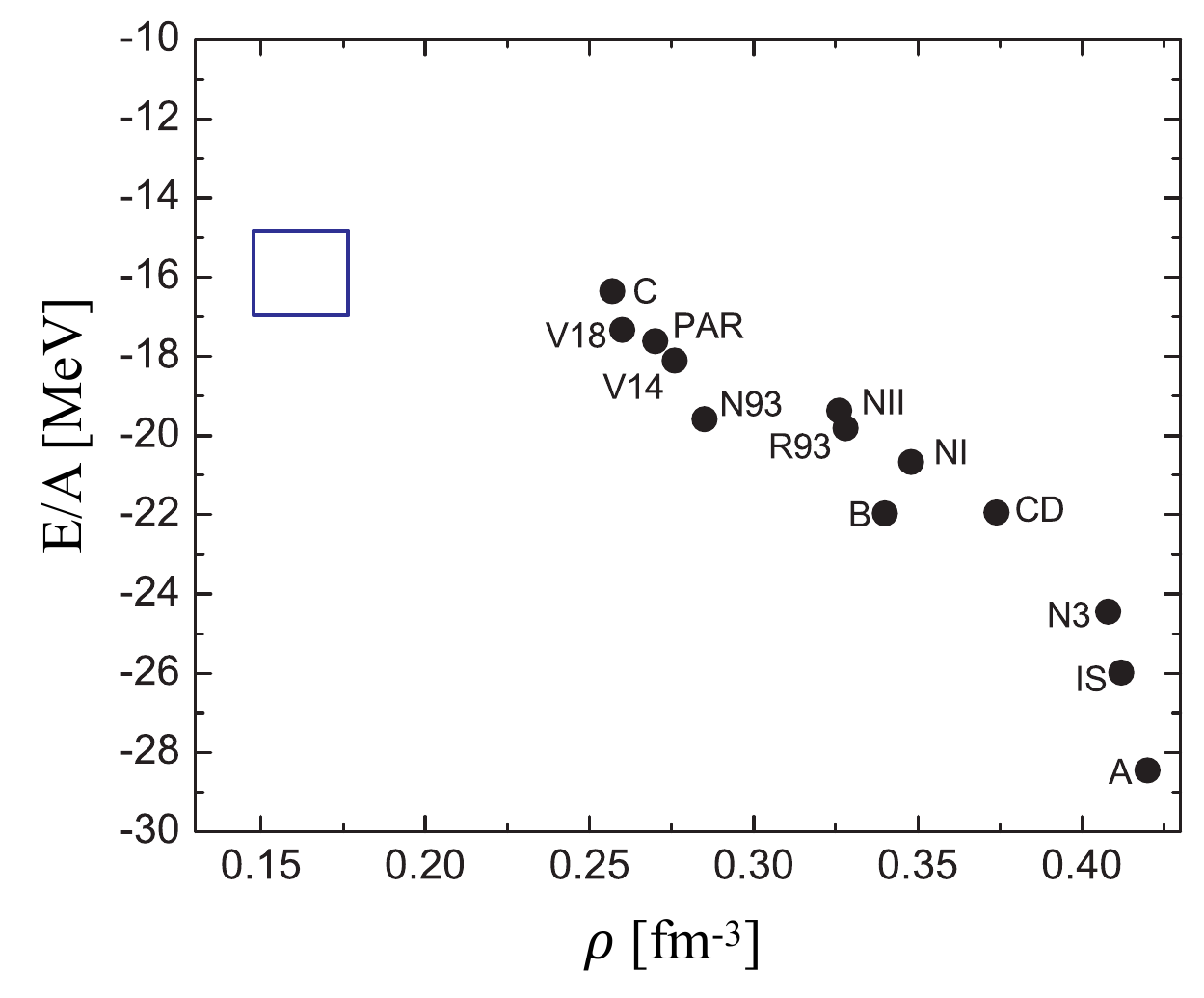}
\end{center}
\caption{Saturation density and energy per particle of isospin-symmetric extended nuclear matter. The blue square indicates the region consistent with known experimental data. Black points represent calculations performed for different NN interactions \textit{without} three-body. The observed correlation in known under the name of Coester line~\cite{Coester70}. Adapted from Ref.~\cite{Li06}.}
\label{fig_coester}
\end{figure}

As calculations on the basis of accurate ($\chi^2 / \text{datum} \approx 1$) OBE two-body potentials became available, systematic deficiencies in the description of both nuclei with $A>2$ and extended nuclear matter emerged.
The former showed typically an underbinding that was increasing with $A$.
In the latter, the saturation point\footnote{I.e. the minimum in the energy per particle as a function of the density.
} was predicted at too high density and energy, see Fig.~\ref{fig_coester}.
It was soon realised that these discrepancies were to be ascribed to the absence of many-body forces in those Hamiltonian models.
As discussed in Sec.~\ref{sec_eff_vs_ab}, the existence of many-nucleon forces is a direct consequence of considering them, effectively, as \textit{structureless} while they do have an internal structure. 
At this point, one might wonder whether all, i.e. up to $A$-body, many-body forces have to be included in a calculation of an $A$-body system.
If not, how many of them are important? 
Chiral effective field theory, discussed in Sec.~\ref{sec_cheft}, suggests that there exists a hierarchy: $x$-body contributions to a given observable in an $A$-body system decrease as $x$ increases.
Phenomenologically, one also observes that the discrepancy with experiment of calculations with OBE two-body potentials (i.e. the part ascribed to many-body forces) is generally much smaller than the interaction energy (i.e. the contribution generated by two-body forces).
The same reasoning can be then repeated (at least in principle) for three- and higher-body forces.

\begin{figure}[t]
\begin{center}
\hspace{-0cm}
\includegraphics[width=7.5cm]{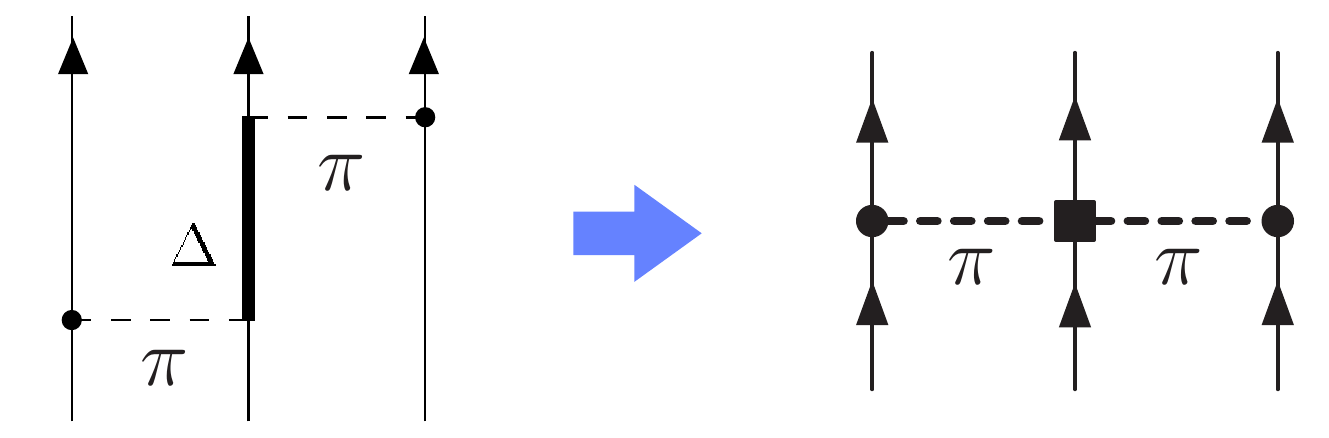}
\end{center}
\caption{Schematic representation of Fujita-Miyazawa three-body force.}
\label{fig_3nf}
\end{figure}

Microscopically, three-body forces account e.g. for processes that involve a nucleon excitation and de-excitation and that can not be described by combining distinct two-body processes. 
A prime example is the $\Delta$ excitation and de-excitation prompted by the exchange of two pions depicted in Fig.~\ref{fig_3nf} and derived already in the 1950s by Fujita and Miyazawa~\cite{fujita57}.
Following the principles of OBE potentials, terms involving other mesons can be constructed and give rise to different operatorial structures in the three-body sector of the Hamiltonian.
Although the importance of individual contributions can be argued about on the bases of physical considerations, a systematic and consistent framework is lacking for OBE potentials such that, in the end, the modelling of three-body operators has relied to a larger extent on phenomenology.

\subsection{Early ab initio calculations}
\label{sec_early_ai}
\begin{figure}[h]
\begin{center}
\hspace{-0cm}
\includegraphics[width=9cm]{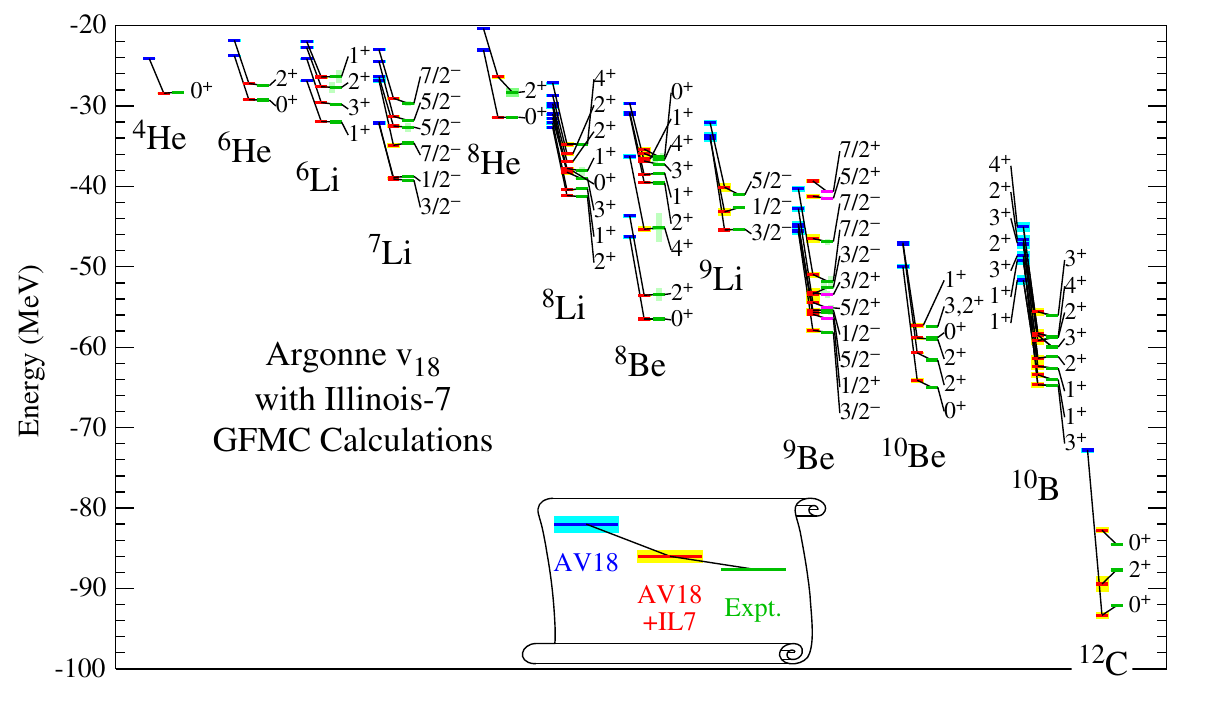}
\end{center}
\caption{Green's function Monte Carlo simulations of ground and first excited states of  light nuclei. Results without and with three-body forces are shown and compared to experimental values. From Ref.~\cite{Carlson15}.}
\label{fig_gfmc}
\end{figure}

In parallel to the modelling of the nuclear Hamiltonian, different many-body methods were developed during the 1980s and 1990s.
The Green's function Monte Carlo (GFMC) approach uses Monte Carlo techniques to sample the many-body wave function in coordinate, spin and isospin space, providing a virtually exact solution of the many-body Schr{\"o}dinger equation~\cite{Carlson15}.
In the 1990s GFMC calculations led the first pioneering set of results in light nuclei~\cite{Pudliner97,Pieper01}, ranging from $^4$He to $^{12}$C, (see Fig.~\ref{fig_gfmc}).
Few years later the no-core shell model (NCSM) approach was put forward and implemented~\cite{Navratil00}. 
As the name indicates, it consists of a valence-space diagonalisation \`a la shell model but without an inert core, i.e. with all active nucleons. 
Analogously to GFMC, it gives a virtually exact solution of the many-body Schr{\"o}dinger equation~\cite{Barrett13}.

These calculations allowed for the first time to make a link between models of basic inter-nucleon interactions and properties of many-nucleon systems. 
In a sense, nuclei were finally simulated ``from scratch".
Nevertheless, such calculations were (and still are) computationally very expensive and were (and, to a certain extent, still are\footnote{A notable exception is constituted by Auxiliary Field Diffusion Monte Carlo (AFDMC) calculations, which have been applied to selected medium-mass nuclei and hypernuclei~\cite{Carlson15,Lonardoni17}}) limited to the light sector of the Segr{\`e} chart.

%%%%%%%%%%%%%%%%%%%%%
\section{Modern models of NN interactions}
\label{sec_new}
%%%%%%%%%%%%%%%%%%%%%

\subsection{Resolution scale}

The development of accurate nucleon-nucleon potentials and the progress in the modelling of three-nucleon forces, together with the formal and computational advances in many-body techniques culminated  - fifty years after the first model of the nuclear force by Yukawa - in the groundbreaking ab initio calculations exemplified in Fig.~\ref{fig_gfmc}.
These results showed that describing the structure properties of (at least light) atomic nuclei starting from models of the basic interactions between their constituents was indeed possible.
However, that was not the end of the story, for two main reasons:
\begin{itemize}
\item[$\circ$] A substantial component of the OBE potentials (in particular three-body forces) remained phenomenological. 
This generates two issues. 
First, the link with the underlying theory, QCD, is lost. 
Second, \textit{predictive power} is not guaranteed, i.e. the fact that known data are well reproduced does not mean that extrapolations to unknown regions of the Segr{\`e} chart are under control.\\
\item[$\circ$] All OBE models featured a strong repulsive short-range component, called ``hard core". 
Physically, it was meant to reflect the impossibility for nucleons to overlap beyond a certain separation, thus avoiding the collapse of nuclear systems.
Mathematically, this hard core induces strong correlations in the nuclear wave function that require the use of sophisticated many-body techniques. 
Practically, this means that large bases are necessary to converge the corresponding ab initio calculations, which are consequently limited to light nuclei.
\end{itemize}
The presence of strong components of the potential at short distances implies, in momentum space, the presence of high momentum components coupled to low-momentum modes.
This entails a description of nuclear systems with very high resolution. 
Is this high resolution really needed?

\begin{figure}[t]
\begin{center}
\hspace{-0cm}
\includegraphics[width=8.5cm]{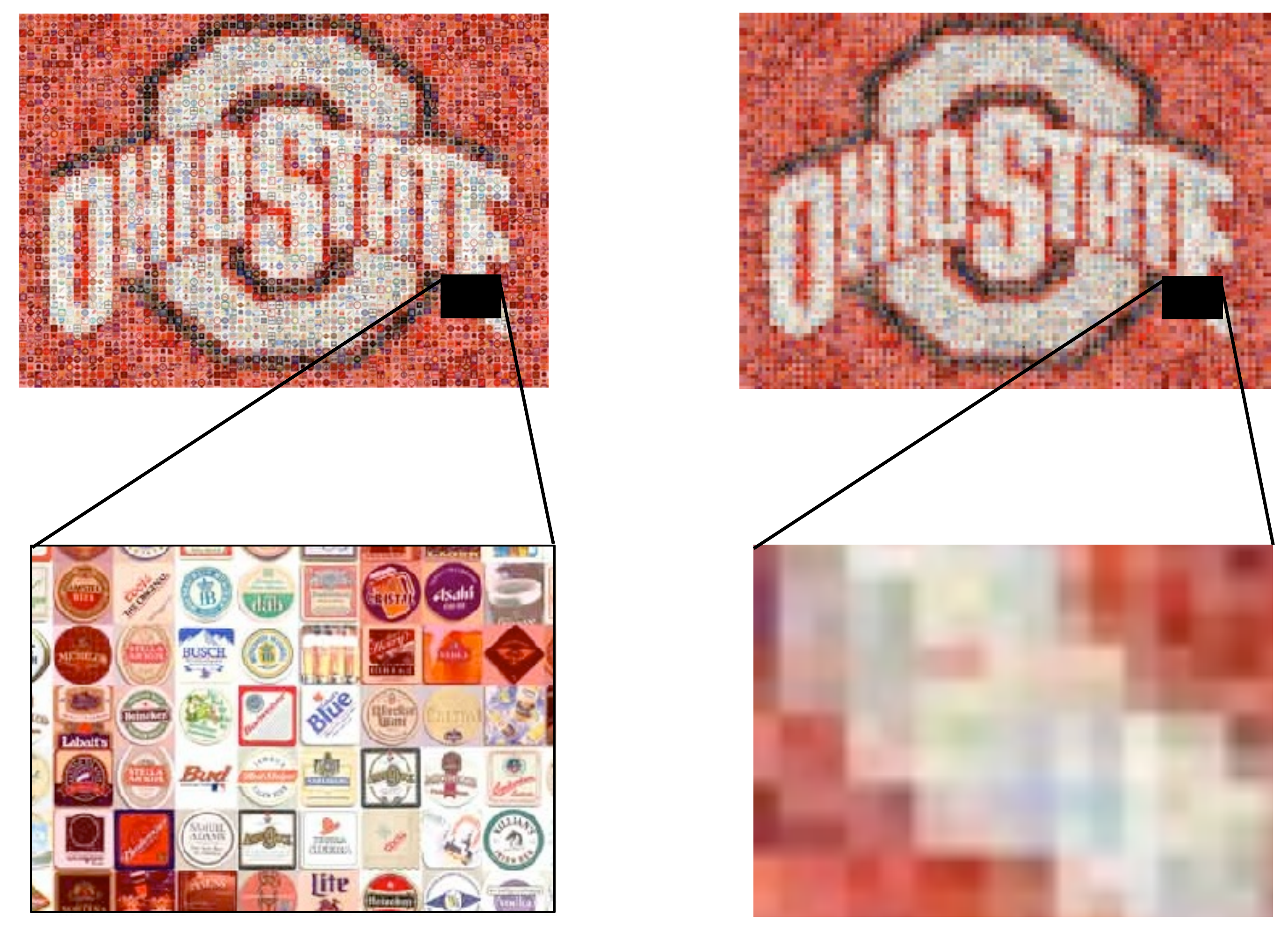}
\end{center}
\caption{Example of high- (left) and low-resolution (right) picture. If we are only interested in reading what is written across the figure, the blurred, i.e. low-resolution, version is as good as the focused, i.e. high-resolution, one. Adapted from~\cite{HebelerOhio}.}
\label{fig_ohio}
\end{figure}

The mesons entering the construction of OBE potentials have masses $> 700$ MeV, which correspond to spatial resolutions $< 0.5$ fm (cf. the nucleon radius $\sim 0.8$ fm). 
This is to be compared with the mass of the pion, $m_\pi \sim 140$ MeV and the average nucleon momenta in nuclei $\sim 200$ MeV.
By contrast, properties of atomic nuclei are associated to low-energy observables that typically range between the keV and the MeV scales\footnote{Although \textit{total} binding energies can reach the GeV for heavy nuclei, the relevant quantity to be compared with the two-nucleon interaction is the energy \textit{per particle}, which is about 8 MeV for nearly all nuclei.}.
Therefore, a mismatch is present between the resolution we impose when constructing our theoretical tools (heavy mesons and hard core) and the object we want to study (low-energy observables). 
This is analogous to looking at a certain object or picture from a distance and not being interested in the tiny details that compose it, as illustrated in Fig.~\ref{fig_ohio}.
Just like in the figure the blurred version is good enough, a description of low-energy observables that does not require resolving high momenta might serve our purpose. 
This does not mean that high-energy physics is disregarded, rather that high-energy details are ``blurred" into some sort of average or effective degrees of freedom.
As a result, the corresponding theory will be simpler and more controllable.

This idea is at the basis of two important breakthroughs that, in the last twenty years, have revolutionised our modelling of nuclear interactions and the way we can perform nuclear structure calculations.
The first, conceptual advance was the application of effective field theory to the construction of nuclear Hamiltonian.
The second, technical advance concerned the application of renormalisation group concepts to post-process these Hamiltonians.

\subsection{Effective field theory and nuclear interactions}
\label{sec_cheft}

Effective field theories (EFTs) are designed according to the following strategy: 
\begin{enumerate}
\item Exploit the separation of (energy or momentum) scales that arises in the physical system to define degrees of freedom and expansion parameter. 
The latter generally takes the form $Q/M$, where $Q$ is the typical momentum at play and $M$ the lowest high-energy scale that is not explicitly included.\\
\item Write all possible terms (e.g. in the Lagrangian) allowed by the symmetries of the underlying theory (QCD in our case).\\
\item Assign a ``degree of importance'' to each term and organise them in a systematic expansion, from the most to the least important.
This procedure is called \textit{power counting} and is at the heart of EFTs.\\
\item Truncate the expansion at a given order and fix the coupling constants using the underlying theory (whenever possible) or  experimental data (our case).\\
\end{enumerate}
The key point resides in the systematic expansion: at each order of truncation, it is expected that the neglected higher orders will contribute \textit{at most} to a certain degree and thus can estimate the error associated to that level of truncation.

\begin{figure}[b]
\begin{center}
\hspace{-0cm}
\includegraphics[width=9cm]{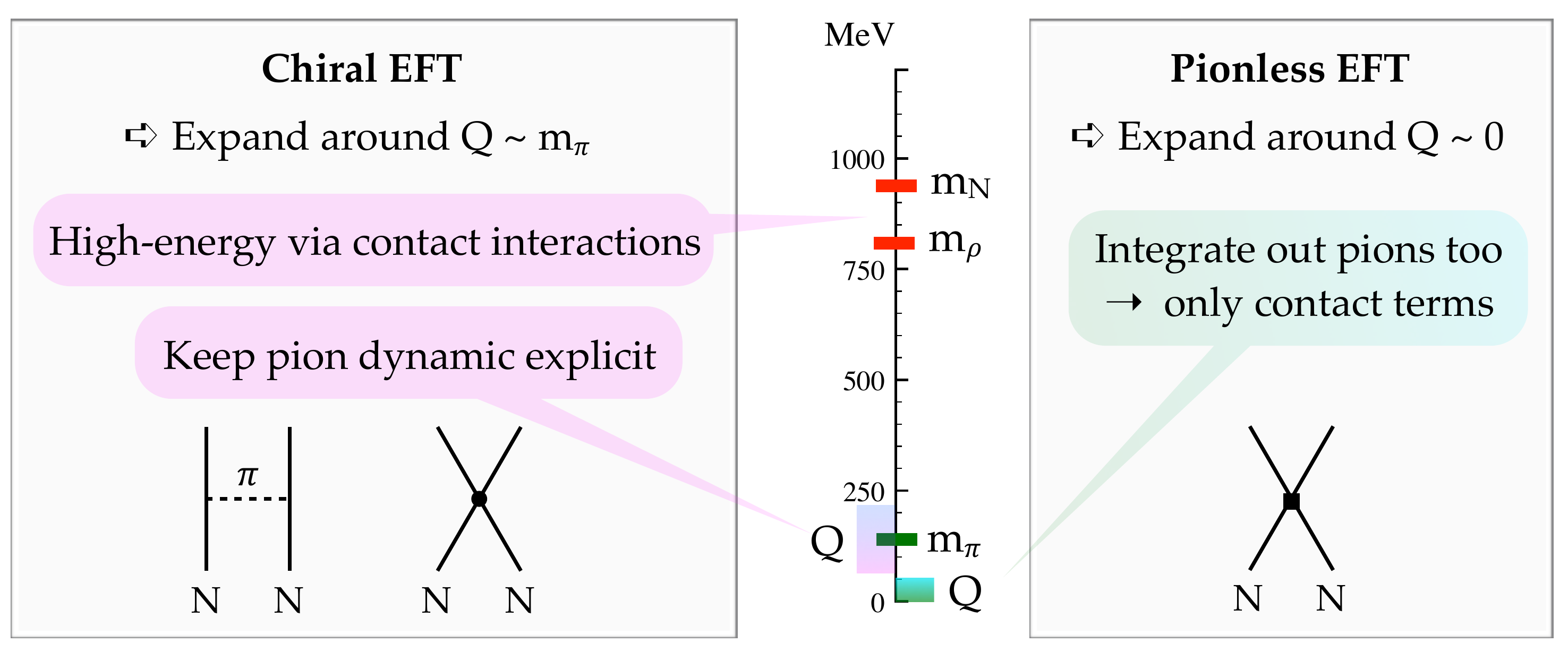}
\end{center}
\caption{Illustration of the two main effective field theories used in low-energy nuclear physics: chiral EFT (left) and pionless EFT (right).}
\label{fig_efts}
\end{figure}
The success of an EFT relies on two important conditions. 
The first one is that it has to be (order-by-order) renormalisable. 
That is, the power counting organisation has to be able to absorb eventual divergencies in the observables at any order in the expansion.
This ensures that any error associated to the use of technical artefacts (e.g. regulator functions that are usually necessary to control diverging integrals) is smaller than the one of the EFT truncation.
The second desirable feature is, clearly, that it reproduces known experimental data (within a few orders in the expansion).
If not, the choice of scale separation and degrees of freedom will have to be revised.

The use of EFTs in nuclear physics was pioneered by Weinberg and collaborators in the early 1990s~\cite{Weinberg90,Weinberg91,vankolck94a,ordonez94} and gave rise to what is known today as \textit{chiral} EFT. 
In chiral EFT interactions between structureless nucleons are modelled via the exchange of pions\footnote{Pions are the pseudo-Goldstone bosons associated to the spontaneous breaking of chiral symmetry in QCD, whence the adjective.} in addition to contact terms that encapsulate higher-energy physics. 
These latter terms take the place of, for instance, what was described through the exchange of heavier mesons in OBE potentials.
The masses of those mesons now correspond to the higher energy scale $M \sim 1$ GeV that is not included explicitly in the EFT.
On the other hand, typical nucleon momenta are $Q \sim 300$  MeV, see Fig.~\ref{fig_efts}.
This yields an expansion parameter of roughly $Q/M \sim 1/3$.

\begin{figure}[t]
\begin{center}
\hspace{-0cm}
\includegraphics[width=8.5cm]{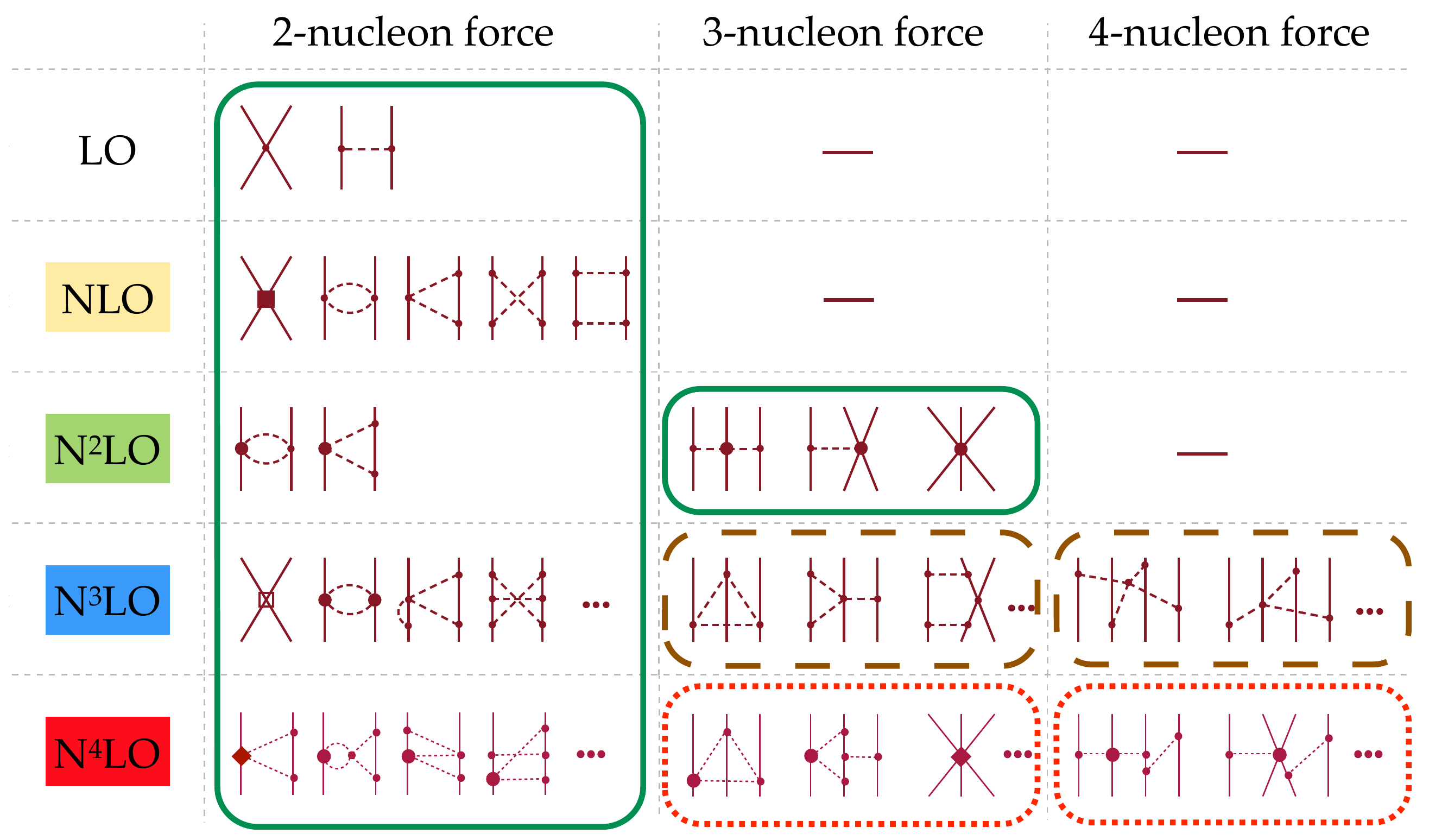}
\end{center}
\caption{Diagrams appearing in the first five orders of chiral EFT derived within Weinberg power counting. 
Dashed lines and dots represent pion exchanges contact interactions respectively.
Sectors contoured with a green solid line have been formally derived and are routinely implemented in nuclear structure calculations. Sectors contoured with a brown dashed line have been formally derived but are not yet routinely implemented in nuclear structure calculations. Sectors contoured with a red dotted line have not been formally derived yet.  The figure has been adapted from Ref.~\cite{Meissner16}, courtesy of Evgeny Epelbaum.}
\label{fig_eft_diagrams}
\end{figure}
\begin{figure}[t]
\begin{center}
\hspace{-0cm}
\includegraphics[width=7.5cm]{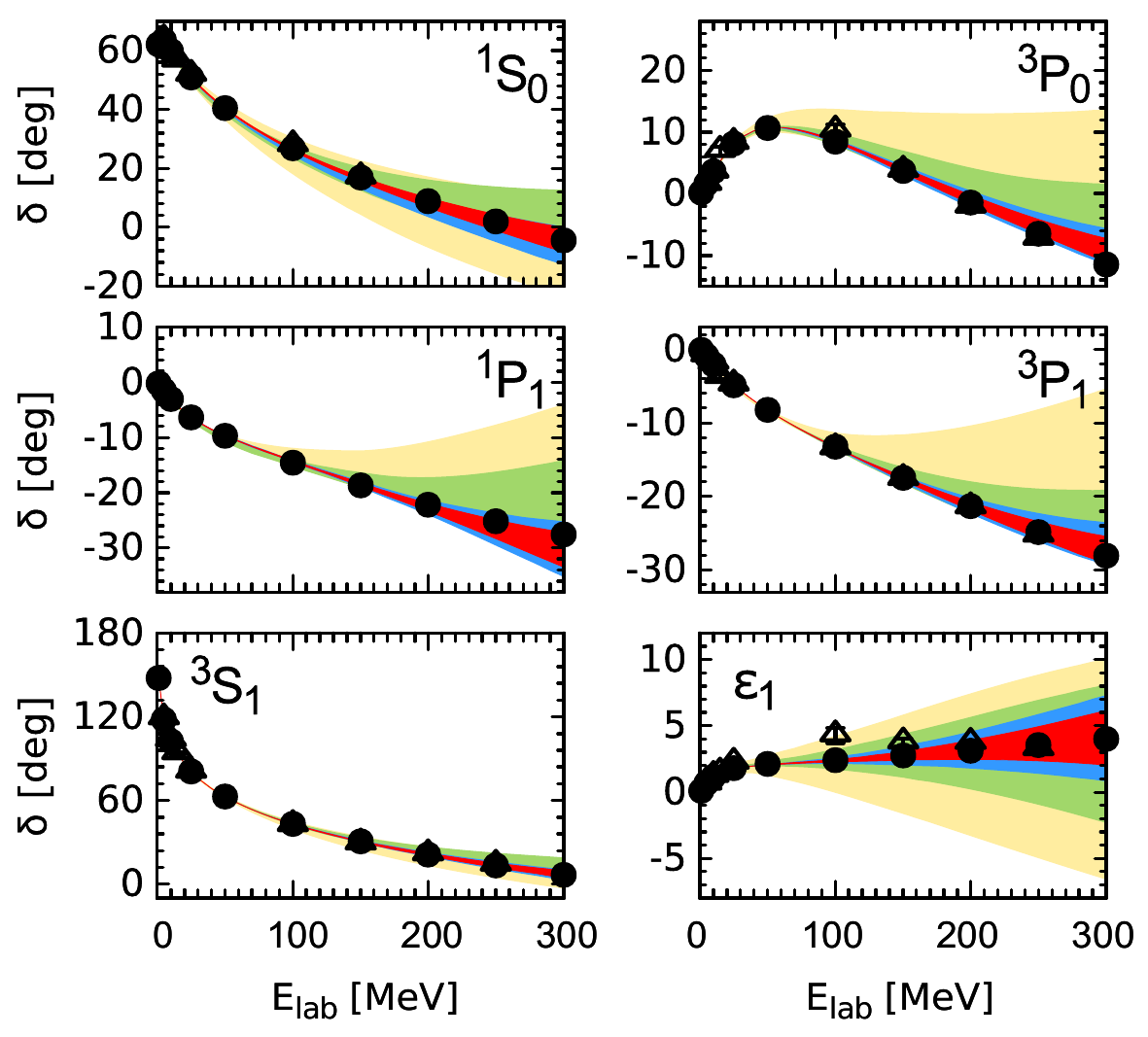}
\end{center}
\caption{Experimental scattering phase shifts (black symbols) vs chiral EFT calculations at different orders. The colour coding is the same as in Fig.~\ref{fig_eft_diagrams}, with different colours referring to different chiral orders. Theoretical error bands are computed following the prescription of Ref.~\cite{Epelbaum14}. The figure has been adapted from Ref.~\cite{Epelbaum15}.}
\label{fig_eft_phase_shifts}
\end{figure}
In Fig.~\ref{fig_eft_diagrams} diagrams entering the potential energy in the first five orders of chiral EFT\footnote{Diagrams shown in Fig.~\ref{fig_eft_diagrams} follow Weinberg power counting rules. For a discussion on different power counting schemes see Section~\ref{sec_th_uncert}.} are displayed.
In particular, one notices that many-nucleon forces naturally arise within this framework and are derived in a consistent fashion.
The first contribution to the three-nucleon sector appears at next-to-next-to-leading order (N$^2$LO), the first four-body forces at N$^3$LO, which corroborates the fact that a natural hierarchy is present among many-body forces.

Starting from the early 2000s, NN interactions constructed within such a scheme begin to reach accuracies comparable to traditional OBE potentials when applied to NN scattering~\cite{Entem03} (see Fig.~\ref{fig_eft_phase_shifts} for a more recent example). 
The modelling of three-nucleon forces followed soon afterwards~\cite{Navratil07}, opening the way to ab initio calculations based on chiral EFT potentials.

Chiral EFT can be seen as an expansion around $Q \sim m_\pi$, where pions have to be necessarily included explicitly.
Nevertheless, if one is interested in the physics at even lower energies, one can imagine going one step further and ``integrating out'' also pions.
The corresponding theory takes the name of \textit{pionless} EFT and can be thought of as an expansion around $Q \sim 0$ (see Fig.~\ref{fig_efts}).
In pionless EFT inter-nucleon forces are modelled solely via contact interactions, which leads to simpler power counting and diagrammatics compared to chiral EFT.
Pionless EFT is typically applied to the description of few-nucleon systems. 
Although it may appear less efficient than the chiral theory when applied to heavier nuclei, its extension is under discussion.

\subsection{Similarity renormalisation group}

Thanks to the implicit and systematic treatment of high-energy degrees of freedom, chiral EFT offers a certain flexibility in modelling the short-range part of the nuclear potential.
In particular, one is not forced to include a strong hard core as in the case of OBE interactions: the resolution scale can be lowered basically ``at will'', with the missing high-energy physics being captured at each scale by the renormalisation of the coupling constants.
In practice, due to presence of two-nucleon bound and resonant states, the potential is derived from the Lagrangian by solving a non-perturbative Lippmann-Schwinger equation.
In order to regulate the loop divergences appearing in the latter, a regulating function with a (momentum) cutoff is usually introduced.
This cutoff sets the corresponding resolution of the potential, as well as its limit of applicability\footnote{Still, the cutoff should not be lowered to the point where relevant physics gets integrated out (in the case of chiral EFT, close to the pion mass).
Clearly, physical quantities, i.e. observables, should not depend on the choice of the regulator, i.e. should be cutoff independent. Cutoff independence should be in principle verified at each order of the EFT expansion.}.

\begin{figure}[t]
\begin{center}
\hspace{-0cm}
\includegraphics[width=8.5cm]{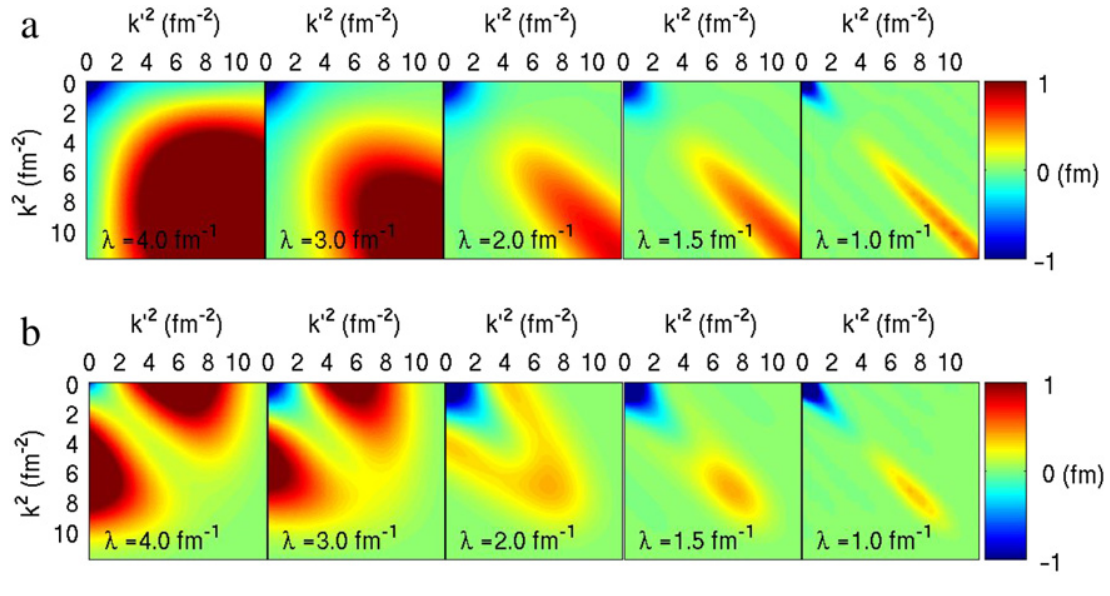}
\end{center}
\caption{Contour plots of momentum-space matrix elements at different stages of a SRG evolution for the $^1$S$_0$ partial wave (a) and the S-wave part of the $^3$S$_1$-$^3$D$_1$ (b). The initial interaction is a N$^3$LO chiral potential with cutoff $\Lambda=600$ MeV~\cite{Entem03}. Taken from Ref.~\cite{Bogner10}}
\label{fig_srg}
\end{figure}

\begin{figure}[t]
\begin{center}
\hspace{-0cm}
\includegraphics[width=8cm]{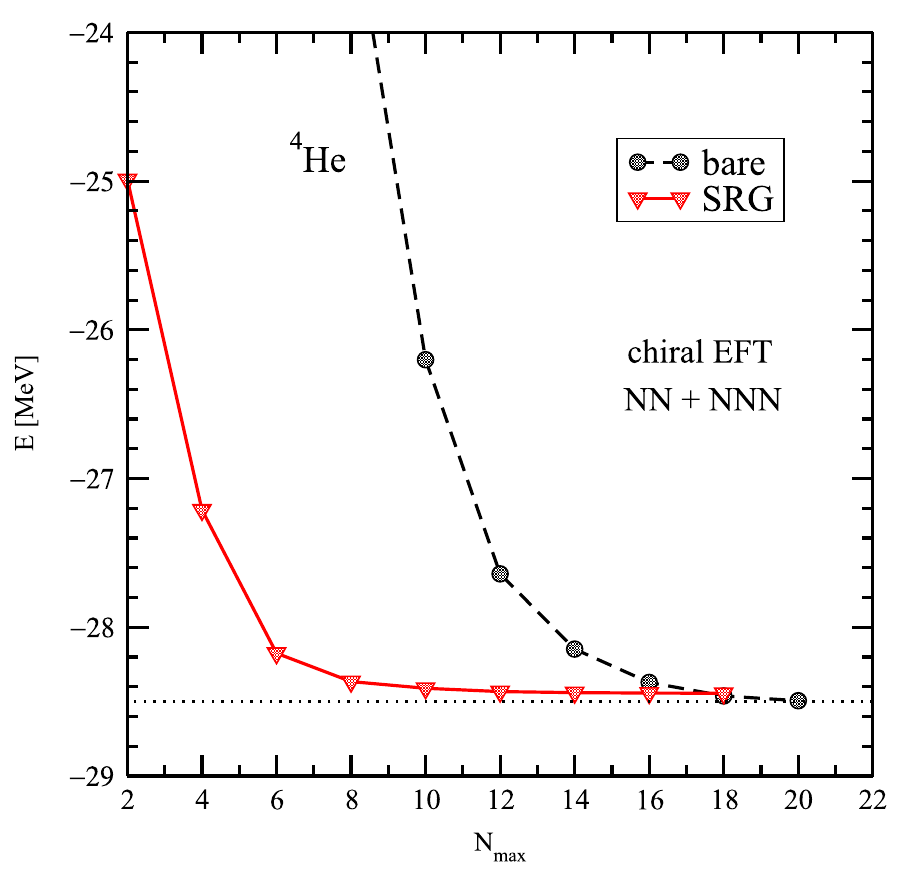}
\end{center}
\caption{No-core shell model calculation of the binding energy of $^4$He with a bare and an SRG-evolved two- plus three-nucleon interaction. Results are displayed as a function of the model space dimension $N_{\text{max}}$.
Taken from Ref.~\cite{Barrett13}.}
% - double check reference
\label{fig_he4}
\end{figure}

Chiral interactions are thus characterised, from the outset, by lower resolution scales compared to OBE potentials.
This allows the use of less sophisticated many-body techniques and smaller basis sets, which enable to enlarge the reach of ab initio simulations.
Can we push this idea further, i.e. can we make couplings between low and high momenta even weaker?
The answer is positive and involves the use of \textit{similarity renormalisation group} (SRG) techniques.
The intuition originates from the fact that any unitary transformation of an operator leaves its eigenvalues unchanged.
Hence, a unitary transformation can be designed to drive the Hamiltonian towards a certain form (e.g. a diagonal form in momentum space), which leads to further decoupling high from low momenta (see Figure~\ref{fig_srg}).
As the SRG evolution drives the interaction towards lower resolution scales, corresponding calculations become easier to converge both in terms of many-body expansion and basis dimension, see Fig.~\ref{fig_he4}.
However, while changing the NN operator, the unitary transformation (which has to be performed in the $A$-body Hilbert space) shifts strength from the two- to the many-body sector.
Therefore, the price to pay is having to deal with three- and perhaps higher-body forces, depending on the SRG evolution.
Since three-body forces have to be included anyway but four-body forces become tricky to handle in practice, a good balance corresponds to an SRG scale that induces three- but not higher-body interactions.

\subsection{Ab initio calculations}
\label{sec_abinitio}

The availability of low-scale potentials, combining EFT and SRG evolution, has revolutionised the field of ab initio simulations of nuclear systems.
While OBE potentials required the use of virtually exact many-body methods of Monte Carlo type, with these modern, soft interactions calculations can be converged at a much lower expense.
In particular, approximate solutions of the full many-body Schr{\"o}dinger equation are already able to provide highly accurate results.

Typically, approximations involve an expansion\footnote{Strictly speaking, there is no well-defined expansion parameter, but the expansion is rather motivated by physical (or phase space, etc.) arguments.} of the exact solution, which is then truncated at some desired degree of accuracy.
The capacity to systematically improve the truncation to ensure that higher orders provide smaller and smaller contributions is therefore crucial\footnote{In the limit of infinite truncation one should obviously recover the exact result.}.

One can identify two types of truncation, one with respect to basis dimensions and the other in terms of many-body correlations.
The first one comes from the fact that the many-body wave function has to be represented over a given (single-particle) basis.
This basis, infinite in principle, has to be finite in practice.
Therefore, one has to show that the employed basis is large enough to contain the relevant physics.
In reference to the previous discussion on resolution scales, the weaker the high-momentum components in the Hamiltonian, the smaller the minimum basis to converge.

In applications to finite nuclei one typically uses the harmonic oscillator (HO) basis, i.e. eigenfunctions of the spherical three-dimensional quantum HO.
Such a basis is suitable to model wave functions of bound nucleons (cf. the shell model discussed in Sec.~\ref{sec_sm}) and possesses several handy analytical properties that bring significant simplifications.
Its drawback resides in the incorrect asymptotic behaviour, which hinders an efficient description of extended wave functions arising e.g. when a nucleus displays a halo structure or when unbound states are considered.
Basis convergence is generally well understood and under control. Also, techniques to extrapolate finite-basis results to infinite basis have been devised~\cite{Furnstahl12,Coon12}.

The starting point to devise a many-body expansion is the definition of a reference state.
This is usually done by splitting the Hamiltonian in two parts, $H = H_0 + H_1$, in such a way that the problem with $H_0$ can be solved exactly.
This solution is then used as a reference around which the exact wave function is expanded.
Again, for a given accuracy, the weaker the high-momentum components in the Hamiltonian, the cruder the approximation one can implement.
A typical reference state in many-body calculations is the solution of a Hartree-Fock problem\footnote{That is, the independent-particle state (Slater determinant) whose energy is the closest to the exact ground-state energy.}.
Expansions are then built in terms of particle-hole excitations on top of the Hartree-Fock state: order after order all the excited configurations contributing to the exact wave function are generated.

One of the simplest expansion techniques is standard many-body perturbation theory. 
Initially abandoned because of the difficulty to treat the repulsive core of early interaction models, it has been revived recently following the advent of low-momentum potentials~\cite{Tichai16}.
More sophisticated, \textit{non-perturbative} approaches have then been devised over the years.
Some of these methods have been initially proposed in the 1950s and 1960s, when many formal developments were carried out.
Nevertheless the first realistic implementations in nuclear structure arrived decades later, and only in the early 2000s modern calculations with microscopic two- and three-body interactions became available.
Examples of these non-perturbative techniques are the self-consistent Green's functions~\cite{Dickhoff04,Cipollone15,Duguet17b} and the coupled cluster approach~\cite{Kowalski04,Hagen14}.
Others, like the in-medium similarity renormalisation group method~\cite{Hergert16a}, have been designed in recent years.
Cross-fertilisation between nuclear physics and quantum chemistry also played an important role for the advance of many-body theory.

\begin{figure}[t]
\begin{center}
\hspace{-0cm}
\includegraphics[width=8.5cm]{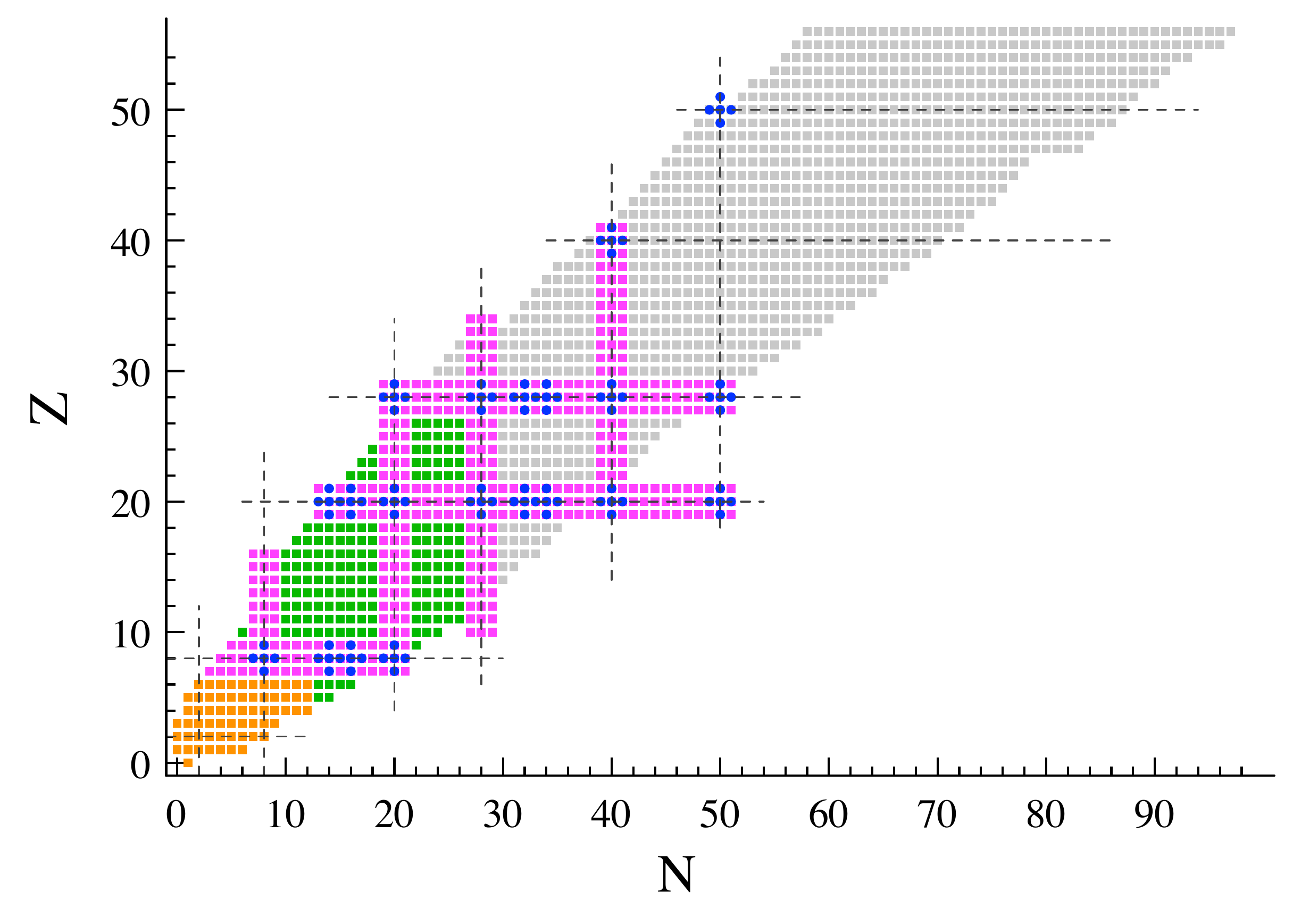}
\end{center}
\caption{Current reach of ab initio calculations. Nuclei in orange represent the area of applicability of ``exact'' techniques like quantum Monte Carlo or no-core shell model. In blue, the reach of approximate methods based on a doubly closed-shell reference state is shown. Such methods have been subsequently generalised to include pairing and thus extended to regions in magenta. Finally, ab initio shell model (green) can be applied between semi-magic chains to cover the remaining medium-mass isotopes.}
% + update figure
\label{fig_aic}
\end{figure}
While exact methods discussed in Sec.~\ref{sec_early_ai} scale exponentially or factorially with the number of particles, approximate methods typically scale polynomially, i.e. much more gently.
This has allowed to extend the reach of ab initio methods from the light sector to the medium-mass region of the Segr{\`e} chart, see Fig.~\ref{fig_aic}.
However, the use of a spherical Hartree-Fock state as reference does not allow an efficient treatment of pairing and/or deformation, which limited the application to doubly-closed shell nuclei.
Starting from 2010, these methods have been expanded to include pairing correlations~\cite{Soma11a,Signoracci15,Hergert17,Tichai18} or even to more general reference states~\cite{Gebrerufael17,Tichai17}.
This allowed to extend their application to full (semi-magic) isotopic or isotonic chains.
Their further generalisation to include deformation would allow calculations in doubly-open shell systems.

Even more recently, some ab initio version of the standard interacting shell model has been designed~\cite{Bogner14,Jansen14}. 
In such an approach, the effective interaction between valence-space nucleons is derived via one of the approximate many-body methods described above. 
By doing so, one maintains the connection to the microscopic Hamiltonian.
Being a full (valence-space) diagonalisation, both pairing and deformation do not constitute an obstacle and the method can be eventually applied between semi-magic chains, complementing the ones based on many-body expansions but working with a spherical reference state.
\begin{figure}[t]
\begin{center}
\hspace{-0cm}
\includegraphics[width=8cm]{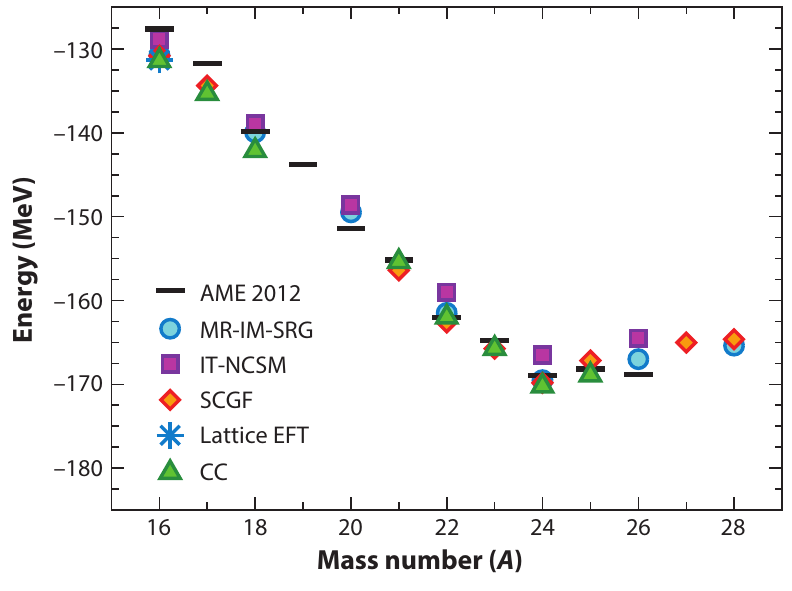}
\end{center}
\caption{Ground-state energies of oxygen isotopes computed with different many-body approaches. Experimental data is shown for comparison. Calculations come from Refs.~\cite{Hergert13,Cipollone13,Epelbaum14,Jansen14}. The compiled figure is taken from Ref.~\cite{Hebeler15}.}
\label{fig_o_benchmark}
\end{figure}

Finally, another method named nuclear lattice EFT has also been developed recently~\cite{Lahde14}.
It combines chiral EFT ideas with quantum Monte Carlo techniques and works in a discretised coordinate box.
Not being tied to the use of a HO basis, the resulting calculations are particularly suitable to describe clusters or other states that are spatially extended.

These methods address the nuclear many-body problem from different perspectives, have sometimes access to different observables and are therefore complementary. 
When they start from the same (two- plus three-body) interaction and target the same observable, the agreement is usually very good. 
A notable example is constituted by the binding energies of oxygen isotopes displayed in Fig.~\ref{fig_o_benchmark}, where different many-body calculations agree within a few percent.
In recent years, this and other successful benchmarks have allowed to pin down uncertainties coming from the many-body expansion and have shifted the spotlight onto those related to the modelling of the nuclear Hamiltonian, which currently dominate total theoretical errors.

%%%%%%%%%%%%%%%%%%%%%
\section{Future challenges}
\label{sec_challenges}
%%%%%%%%%%%%%%%%%%%%%

% + bridge structure and reactions

\subsection{Towards thorough theoretical uncertainties}
\label{sec_th_uncert}

Current ab initio calculations rely on a double expansion: on the one hand the EFT expansion yielding the nuclear potential at different orders, on the other the many-body expansion building more and more sophisticated approximations to the exact solution.
This scheme is systematically improvable in the sense that as one includes higher and higher orders (in either expansion) one approaches the exact result.
Ideally, at each order of truncation (in either expansion), one should be able to associate an error accounting for the missing (higher-order) terms. 
For what concerns many-body truncation errors, this can be determined by either estimating the contribution of the next order or comparing to exact methods whenever possible.
Current many-body truncations have proved to be sufficiently accurate, with corresponding errors of a few percent.
This estimate is consistent e.g. with the spread between different calculations shown in Fig.~\ref{fig_o_benchmark}.

The situation is more delicate for the EFT expansion.
In principle, one of the advantages of using an effective theory is indeed the capacity to associate an error to each level of truncation of the theory. 
In practice, the way chiral EFT is currently implemented poses questions about its feasibility.
A fundamental issue concerns the viability of Weinberg power counting, at the basis of modern chiral EFT interactions, with its correctness being debated (see e.g.~\cite{Long16}).
Alternative power counting formulations have been proposed but not yet exploited to construct full Hamiltonians.
Moreover, a practical issue relates to the difficulty of deriving higher orders in the chiral EFT expansion and translating them into matrix elements usable by many-body practitioners, which hinders order-by-order many-body calculations.
Nevertheless, progress is being made towards the long-term goal of thoroughly assessing associated errors and propagating them into the calculation of many-body observables.

\subsection{Extending ab initio calculations to heavy nuclei}

Provided that a suitable interaction model is at hand, current ab initio implementations are limited in their applicability to around mass $A \sim 100$.
The reasons are mainly computational, but formal challenges are present as well. 
For what concerns shell model-type calculations, a diagonalisation of the valence-space Hamiltonian is involved.
As $A$ increases, the dimension of the needed valence space increases.
Around or slightly above $A \sim 100$ the number of matrix elements associated to those valence spaces hits the limits of aggregate memory available in modern high-performance computing clusters (see Fig.~\ref{fig_diag_dim}).
Possible solutions involve the use of importance-truncation techniques to pre-select a subset of matrix elements that enter the diagonalisation~\cite{Roth09} or the use of Monte Carlo methods~\cite{Shimizu12}.

Expansion methods face a different computational problem as they require the use, i.e. the computation and storage, of large tensors.
This pertains both to the interaction matrix elements, in particular of three-body operators, and to the (particle-hole) amplitudes that enter the many-body expansion.
As the mass and consequently the basis increases, these tensors become intractable. 
A possible solution involves the implementation of tensor-decomposition techniques developed in applied mathematics, already in use in quantum chemistry~\cite{Schutski17}.
In addition, these many-body approaches require generalisations to address doubly open-shell systems, where collective correlations - difficult to capture when the expansion builds on a spherical reference state - become significant. 
First steps in this direction are being done~\cite{Yao18}.

In general, the extension of ab initio calculations to heavy nuclei will necessarily involve significant technical and computational developments.
Even if such calculations might be able to cover, one day, the whole nuclear chart, at present it is not clear whether this will be the preferable strategy for a predictive, universal first-principle approach or instead other EFTs, e.g. based on different (more collective) degrees of freedom, will turn out to be more efficient~\cite{ESNT17}.
\begin{figure}[b]
\begin{center}
\hspace{-0cm}
\includegraphics[width=8.5cm]{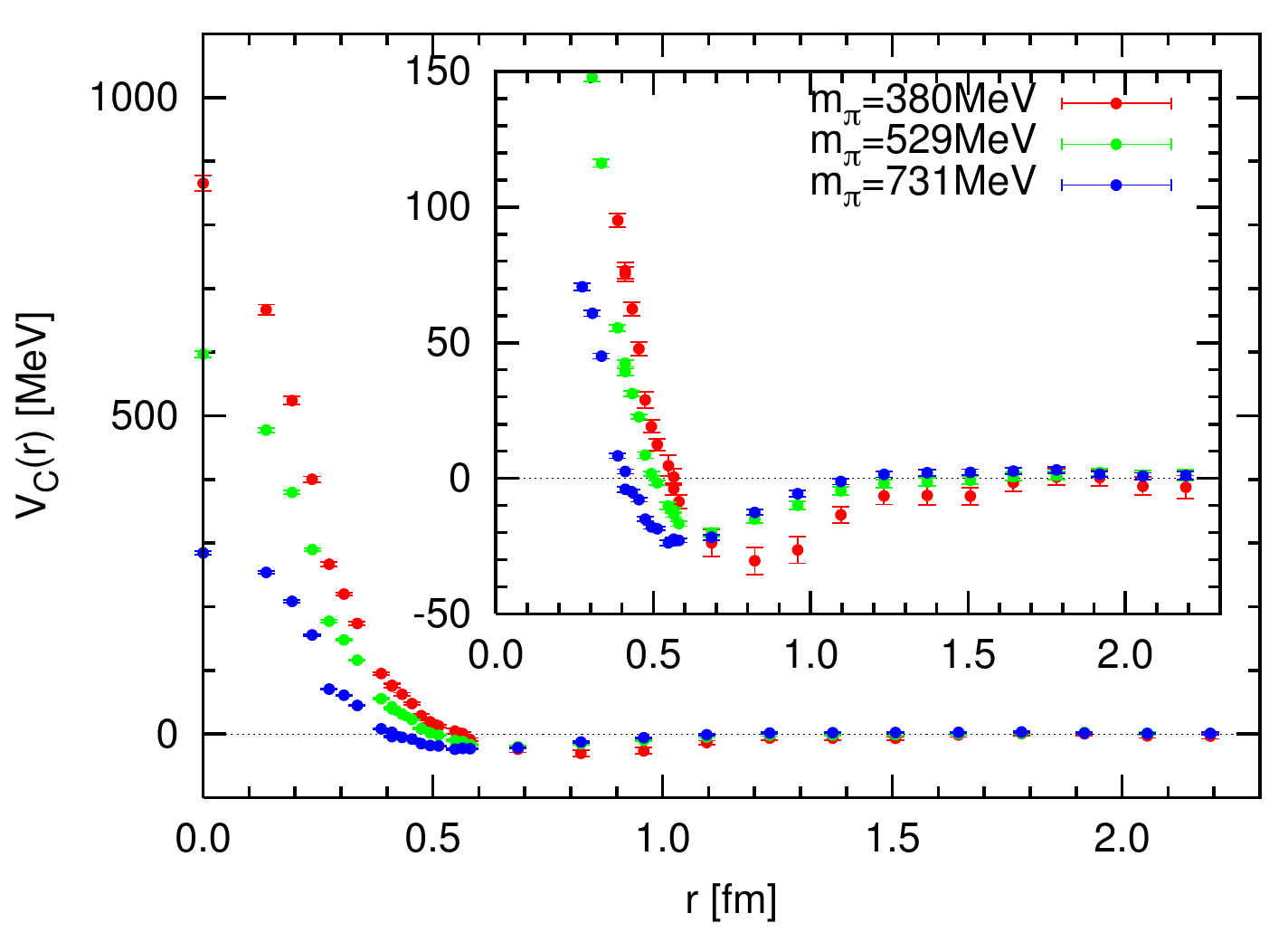}
\end{center}
\caption{Central part of the nucleon-nucleon potential in the $^1$S$_0$ channel computed within lattice QCD for three different quark masses. Taken from Ref.~\cite{Ishii07b}.}
\label{fig_qcd_pots}
\end{figure}
\begin{figure}[b]
\begin{center}
\hspace{-0cm}
\includegraphics[width=8cm]{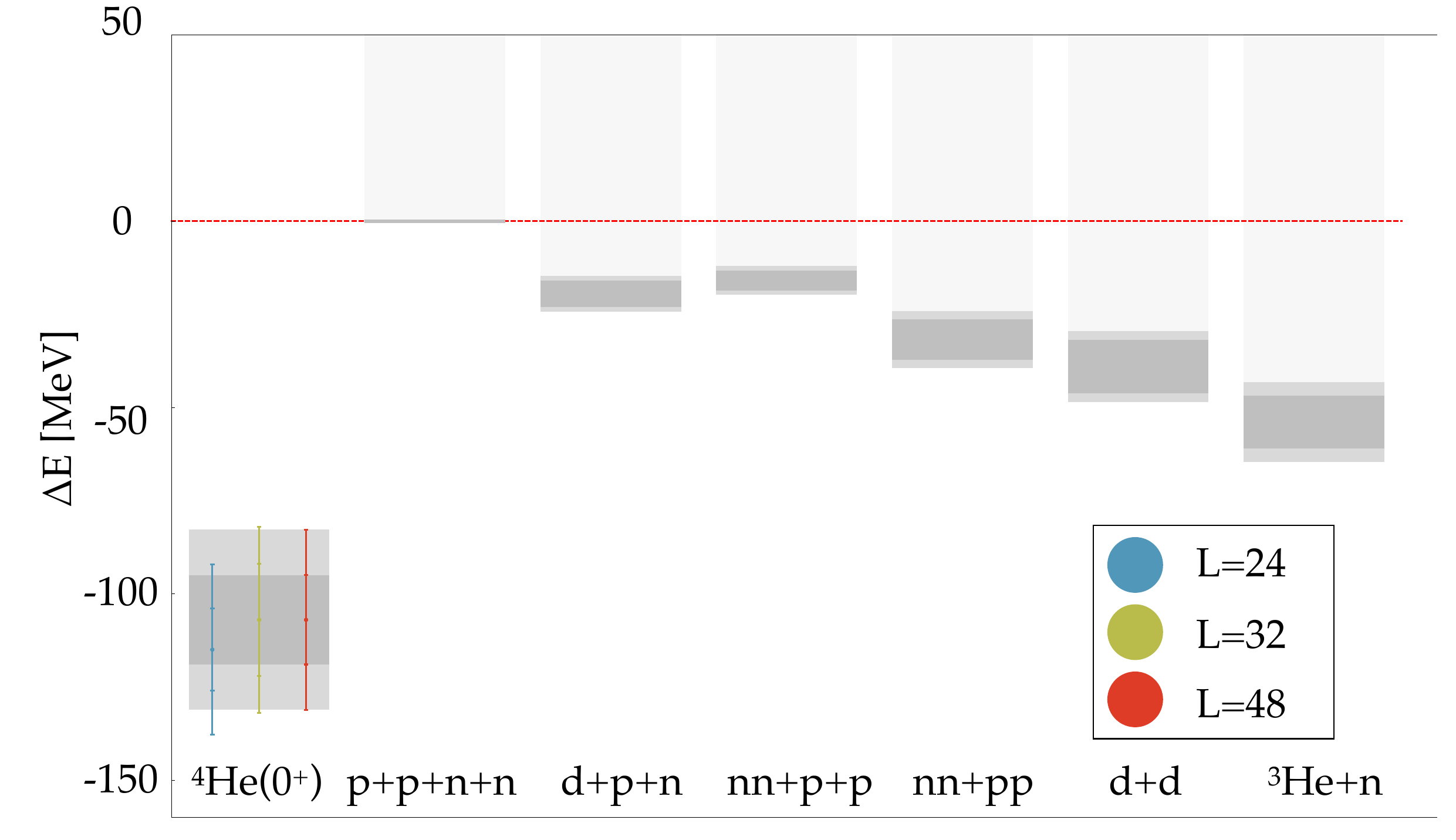}
\end{center}
\caption{Lattice QCD calculations of bound-state energy levels in the $^4$He sector. Adapted from Ref.~\cite{Beane12}.}
\label{fig_qcd_masses}
\end{figure}

\subsection{Lattice QCD}

One could argue that working with nucleons and pions as degrees of freedom is not really ``ab initio'', since we know that they are composite particles governed by the underlying theory of quantum chromodynamics. 
Then, can we compute properties of atomic nuclei starting from QCD? 

At low energy, QCD is non-perturbative and calculations are possible only via lattice simulations.
A possibility consists in constructing the bare nucleon-nucleon (and higher-body) interaction directly from lattice QCD calculations.
This route is being pursued but, although a two-body potential has been successfully computed~\cite{Ishii07a} (see Fig.~\ref{fig_qcd_pots}) and even applied to compute properties of light nuclei~\cite{McIlroy18}, considerable difficulties remain in the three-nucleon sector.
In addition, the extraction of such potential is model-dependent and might lead to further issues.
Alternatively, one could envisage to simulate directly nucleon-nucleon scattering phase shifts and then use them to fit a given interaction model, e.g. from chiral EFT.
This strategy could be advantageous in those channels where experimental information is missing or hard to collect.

Finally, lattice calculations of hadron masses have resulted very successful over the last few years.
What about simulating \textit{directly} multi-baryon systems like atomic nuclei?
After a period of development, the first lattice QCD calculations of nuclear masses are indeed being worked out~\cite{Beane12}.
As evident from Fig.~\ref{fig_qcd_masses}, these results are still far from producing realistic predictions.
The main issues relate to the use of a pion mass that often does not coincide with the physical one and the need for high statistics since nuclear excitations are typically much smaller than QCD scales.
At this point, the ambition to cover the whole Segr{\`e} chart seem to require arduous efforts and probably looks illusory - but time will tell us.

\section*{Acknowledgements}

I thank Angela Bonaccorso and the other organisers of the Summer School ``\textit{Rewriting Nuclear Physics textbooks: basic nuclear interactions and their link to nuclear processes in the cosmos and on earth}'' at the University of Pisa for the opportunity of giving a lecture and writing the present article.
I also thank P. Arthuis, T. Duguet and A. Tichai for useful comments on the paper.

\bibliographystyle{epj} 
\bibliography{biblio}

\end{document}